\documentclass[11pt,a4paper]{extarticle}
\usepackage[T1]{fontenc}
\usepackage[latin9]{inputenc}
\usepackage{array}
\usepackage{verbatim}
\usepackage{float}
\usepackage{units}
\usepackage{amsmath}
\usepackage{amssymb}
\usepackage{graphicx}
\usepackage{wasysym}
\usepackage{esint}

\makeatletter

\providecommand{\tabularnewline}{\\}

\usepackage[english]{babel}
\pdfoutput=1 % if your are submitting a pdflatex (i.e. if you have
\usepackage{jheppub}% for details on the use of the package, please
\title{\boldmath Probing compressed mass spectra in electroweak supersymmetry with Recursive Jigsaw Reconstruction}

\author[a,b]{M. Santoni}%,\note{Corresponding author.}}

\affiliation[a]{University of Adelaide, Department of Physics, Adelaide, SA 5005, Australia}
\affiliation[b]{ARC Centre of Excellence for Particle Physics at the Tera-scale, University of Adelaide}

\emailAdd{marco.santoni@adelaide.edu.au}

\abstract{The lack of evidence for the production of colored supersymmetric particles at the LHC has increased interest in searches for superpartners of the electroweak SM gauge bosons, namely the neutralinos and charginos. These are challenging due to the weak nature of the production process, and the existing discovery reach has significant gaps in due to the difficulty of separating the supersymmetric signal from SM diboson events that produce similar final states and kinematics. We apply the Recursive Jigsaw Reconstruction technique to study final states enriched in charged leptons and missing transverse momentum, focusing on compressed topologies with direct production of charginos and neutralinos decaying to the lightest neutral supersymmetric particle through the emission of $W$ and $Z$ bosons. After presenting prototype analysis designs for future LHC runs, we demonstrate that its detectors have the potential to probe a significant amount of unexplored parameter space for chargino-neutralino associated production within the next few years, and show that the very challenging successful search for chargino pair production with compressed spectra might be possible by the end of the LHC lifetime.
}

\@ifundefined{showcaptionsetup}{}{%
 \PassOptionsToPackage{caption=false}{subfig}}
\usepackage{subfig}
\makeatother

\begin{document}
\maketitle \flushbottom

\section{Introduction }

The Standard Model (SM) of particle physics provides a successful
explanation for a multitude of phenomena, but is considered an effective
field theory of an underlying model valid at higher energies. Supersymmetry (SUSY)
\cite{susyATLAS1Miyazawa:1966mfa,susyATLAS2Ramond:1971gb,Susyatlas3Golfand:1971iw,susyatlas4Neveu:1971rx,susyatlas5Neveu:1971iv,susyatlas7Volkov:1973ix,Wess197452,Wess:1974tw}
refers to an invariance under generalized space-time transformations
linking bosons and fermions. For each Standard Model particle it is
postulated that there exists a partner with spin differing by one-half
unit and other quantum numbers unchanged. Evidence of supersymmetric
particles with mass at the electroweak-TeV scale are a sought-after
experimental outcome, providing an explanation for the stabilization
of the Higgs mass and potential gauge coupling unification. 

R-parity conserving SUSY phenomenology demands an even number of superparticles
in each interaction. Consequently, at a collider experiment, super-partners
would be pair-produced, providing in the final state two stable lightest
supersymmetric particles (LSPs). As the LSP interacts only weakly,
escaping the detector, it assumes the characteristics compatible with
a dark matter candidate. At the Large Hadron Collider (LHC) the missing
transverse momentum ($\vec{{\not\mathrel{E}}}_{T}$) can infer the
presence of unmeasured weakly interacting particles. As the momentum
in the transverse plane is conserved any missing momentum can be assumed
to arise from missing particles including the LSPs.

Reconstruction techniques suffer from the lack of information related
to the multiplicity and the masses of the particles not interacting
with the detector. The difficulty is exacerbated in situations where
the momenta of the weakly interacting particles is low. This is the
case for supersymmetric spectra in which the difference in mass between
the pair-produced parent superparticles $\tilde{P}$ and the LSPs
is low. Scenarios with a small mass-splitting are referred to herein
as \textit{compressed} \cite{An:2015uwa}. 

In the compressed regime, visible and invisible decay products have
low transverse momenta, as the center-of-mass system of the parent
superparticles does. In this scenario, the efficacy of typical variables
\cite{Hinchliffe:1996iu} exploited to distinguish signal from background,
based on large object transverse momenta and missing transverse energy,
is limited. 

One can gain indirect sensitivity by observing the reaction of the
LSP pair to a probing force. The initial state radiation (ISR) from
the interacting partons is the natural probe provided in the laboratory
of a hadron collider. The ISR can boost the sparticles produced in these 
reactions and in turn endow their decay products with its momentum.

The Recursive Jigsaw Reconstruction technique has been used by the ATLAS collaboration 
to probe supersymmetric scenarios in cases involving the production
of colored superpartners of SM particles \cite{ATLAS-CONF-2017-022}. 
In the compressed regime, a general basis of kinematic observables designed for the analysis of
events with initial state radiation can be used independently from
the topology investigated. In this paper we focus on applications
of this technique to final states in the electroweak SUSY sector.

Electroweakinos are a linear combination of the fermionic partner
of the gauge bosons and the two Higgs bosons. Neutral higgsinos and
gauginos mix to form four eigenstates of mass called neutralinos ($\tilde{\chi}_{i}^{0}$
with $i=1,2$, 3 or 4) while charged winos and higgsinos form two
eigenstates of mass referred to charginos ($\tilde{\chi}_{i}^{\pm}$
with $i=1$ or 2). Herein one assumes the lightest neutralino is
the LSP and focuses on cases with small mass difference between it
and the $\tilde{\chi}_{1}^{\pm}$ and/or $\tilde{\chi}_{2}^{0}$.
Compressed scenarios involving electroweakinos are common in supersymmetric
theory. For example, in naturalness-inspired models \cite{Barbieri:1987fn}
the higgsino components are light, hence small mass splittings are expected
for the lower eigenstates of mass of higgsino-like charginos and neutralinos.
At the same time the masses of the wino components, appearing in the
one-loop corrections to the Higgs mass, are expected to be limited. 

Qualitatively, the smaller the mass difference between the parent
superparticle and the LSP, $\Delta M=M_{\tilde{P}}-M_{\tilde{\chi}_{1}^{0}},$
the less opportunity there is to accommodate an additional intermediate
superparticle. In this study, final states arising from intermediate
sleptons are not considered. We focus on simplified topologies where
$\tilde{\chi}_{1}^{\pm}\rightarrow W^{\pm}\tilde{\chi}_{1}^{0}$ and
$\tilde{\chi}_{2}^{0}\rightarrow Z\tilde{\chi}_{1}^{0}$. For a mass
splitting below the $W$-boson mass, two-body decays are kinematically
suppressed and we generate three-body decays involving off-shell bosons
and assume other mediators do not contribute.

\section{The simplified compressed decay tree}
Recursive Jigsaw Reconstruction (RJR) \cite{Jackson:2017gcy} is a HEP technique 
based on the imposition of 
a decay tree mimicking the signal topology investigated. A series of algorithms, 
referred to as jigsaw rules, are applied to solve final state ambiguities due to unknown kinematic 
degrees of freedom when weakly interacting particles are present and 
combinatoric challenges due to the presence of indistinguishable visible particles 
from a detector prospective. The result is an estimate of the relevant reference frames 
and hence, a complete basis of kinematic observables sensitive to 
the masses and decay angles of the resonances appearing in the chosen tree, 
which can be used to distinguish signatures of new physics from the SM background. 

\begin{figure}[t]
\noindent \centering{}\includegraphics[height=5cm]{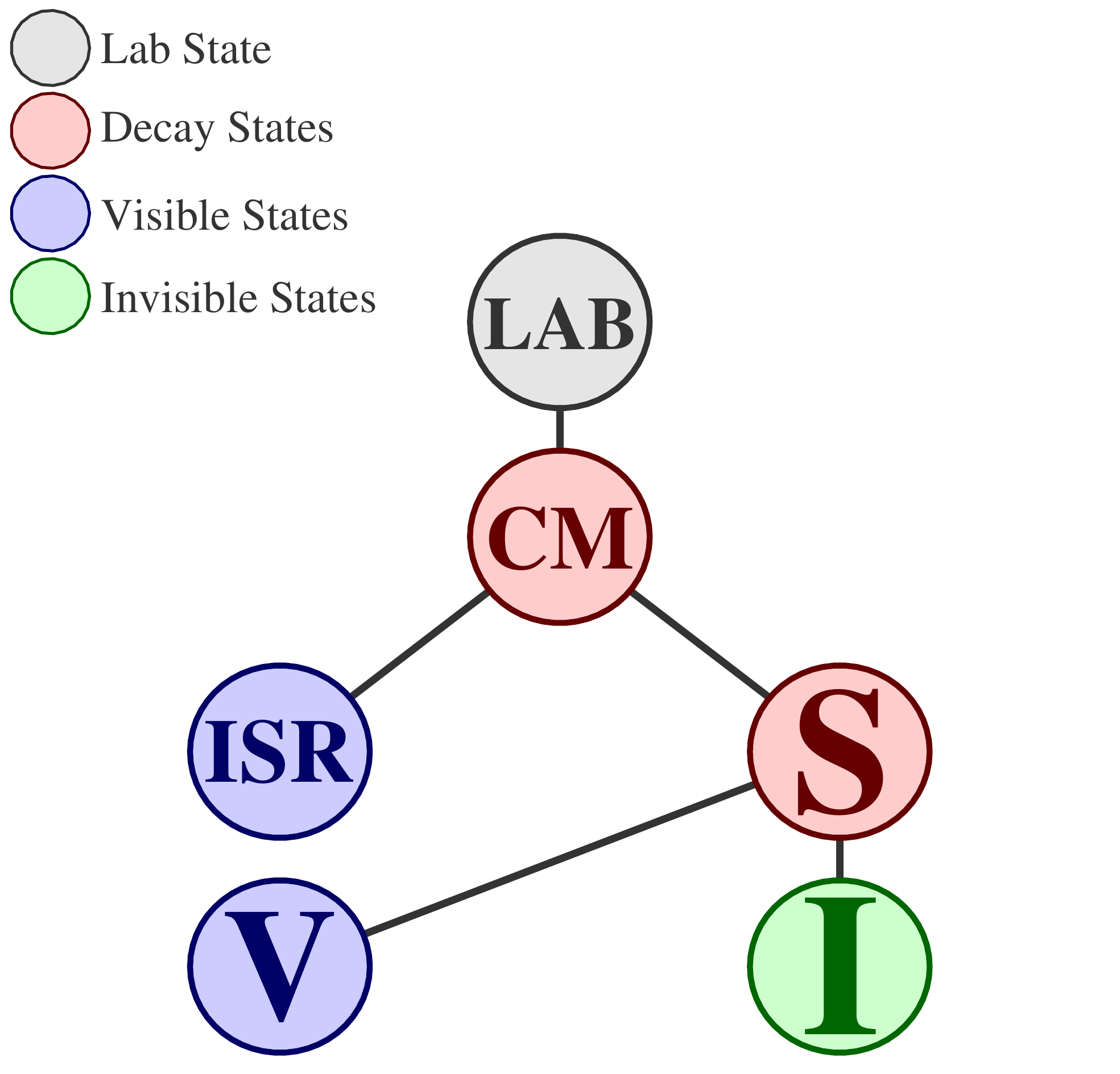}\caption{The simplified decay tree diagram for the analyses of compressed signal
topologies.\label{fig:The-compressed-decay}}
\end{figure}

In RJR involving compressed scenarios the
simplified decay tree shown in Figure \ref{fig:The-compressed-decay}
is used for analyzing topologies with initial state radiation. 
A transverse view of the event is considered,
namely all the $z$-momenta of the visible objects are set to zero. 

We follow the procedure outlined in \cite{PhysRevD.95.035031}.
The estimate for the center-of-mass system of the whole reaction
SUSY + ISR is labeled by CM, ISR is the system assigned to the radiation
from the initial state, S is the signal or sparticle system decaying
in visible and invisible products and hence to the V and I systems.
In each event, the missing transverse momentum is assigned to the
I-system, while a jigsaw rule specifies the reconstructed objects
hypothesized to come from the decay of sparticles and assigned to
the V-system with respect to those associated with ISR.

Topology independent observables include:
\begin{itemize}
\item $R_{\mathrm{ISR}}\equiv\frac{\left|\vec{p}_{\mathrm{I},T}^{\mathrm{\,CM}}\cdot\hat{p}_{\mathrm{ISR},T}^{\mathrm{CM}}\right|}{p_{\mathrm{ISR},T}^{\mathrm{CM}}}$:
variable sensitive to the mass ratio between LSP and parent superparticle.
\item $p_{\mathrm{ISR},T}^{\mathrm{CM}}$: magnitude of the vector-sum of the
jets transverse momenta of the ISR-system evaluated in the CM frame.
\item $\Delta\phi_{\mathrm{ISR},\mathrm{I}}$: opening angle between the
ISR-system and the I-system, evaluated in the CM frame.
\end{itemize}
In final states with two LSPs and no other weakly interacting particles
the observable $R_{\mathrm{ISR}}$ can be written in the laboratory
frame as

\begin{equation}
R_{\mathrm{ISR}}\sim\frac{\left|\vec{{\not\mathrel{E}}}_{T}\cdot\hat{p}_{T}^{ISR}\right|}{p_{T}^{ISR}}\sim\frac{M_{\tilde{\chi}_{1}^{0}}}{M_{\tilde{P}}}+\mathcal{O}\left(\frac{p_{\tilde{\chi}_{1}^{0}}^{\tilde{P}}}{2M_{\tilde{P}}}\right)\left(\frac{\sqrt{\left(p_{T}^{ISR}\right)^{2}+m_{\tilde{P}\tilde{P}}^{2}}}{p_{T}^{ISR}}\right)\sin\Omega .\label{eq:RISRfull}
\end{equation}
This approximation is valid for the extreme compressed scenarios,
hence in the limit of a low-momentum of the LSP in the parent superparticle rest
frame $p_{\tilde{\chi}_{1}^{0}}^{\tilde{P}}$ with respect to the parent superparticle
mass $M_{\tilde{P}}$, while $m_{\tilde{P}\tilde{P}}$ is the true mass of
the S-system and $\sin\Omega$ is a quantity which is zero on average.
The observable scales with the mass ratio $\nicefrac{M_{\tilde{\chi}_{1}^{0}}}{M_{\tilde{P}}}$
and width of order $\nicefrac{p_{\tilde{\chi}_{1}^{0}}^{\tilde{P}}}{2M_{\tilde{P}}}$
in the limit $p_{T}^{ISR}\gg m_{\tilde{P}\tilde{P}}$. When visible decay
objects are not reconstructed in the V-system or additional neutrinos
in the final state contribute to the missing transverse momentum,
$R_{\mathrm{ISR}}$ is expected to assume values between the mass
ratio and one with smaller resolution.

\section{Compressed electroweakino production in leptonic channels}

Simulated Monte Carlo (MC) samples of Standard Model backgrounds and SUSY 
signals are used to study distributions of the performance of the RJR observables.
The SM background processes expected to be the largest contributions have been generated
elsewhere \cite{SnowmassEnergysimulationsAnderson:2013kxz}.
These samples are proton proton collisions at $\sqrt{s}=14$ TeV generated
with MadGraph 5 \cite{Alwall:2011uj}. The parton shower and hadronization
is performed with Pythia 6 \cite{Sjostrand:2006za} followed by a
detailed detector simulation with Delphes 3 \cite{Selvaggi:2014mya}
in which a parameterization for the performance of the existing ATLAS
\cite{Aad:2008zzm} and CMS \cite{Chatrchyan:2008aa} experiments
is implemented. 
Jets are reconstructed by the anti-$k_{T}$ clustering algorithm \cite{Cacciari2008} 
with $R=0.5$ and $p_{T}^\mathrm{{min}}=20$ GeV, 
implemented with the FastJet \cite{Cacciari2012} package.
The simulation procedure involves generation of events
at leading order in bins of the scalar sum of the generator level
particles transverse momenta, with jet-parton matching and corrections
for next-to-leading order (NLO) contributions \cite{Snowmass:2013onh}.

The same procedure and parametrization are used to generate the signal
samples. The topologies considered are associated chargino-neutralino
production and chargino pair production assuming degenerate masses
$M_{\tilde{\chi}_{1}^{\pm}}=M_{\tilde{\chi}_{2}^{0}},$ $M_{\tilde{\chi}_{1}^{+}}=M_{\tilde{\chi}_{1}^{-}}$
in the range 100 GeV$\leq M_{\tilde{P}}\leq$ 500 GeV in the compressed
regime: the mass splittings considered are in the range 15 GeV$\leq\Delta M\leq75$
GeV. 

\begin{figure}[t]
\noindent \centering{}\subfloat[\label{fig:cc}]{\noindent \begin{centering}
\includegraphics[height=3.2cm]{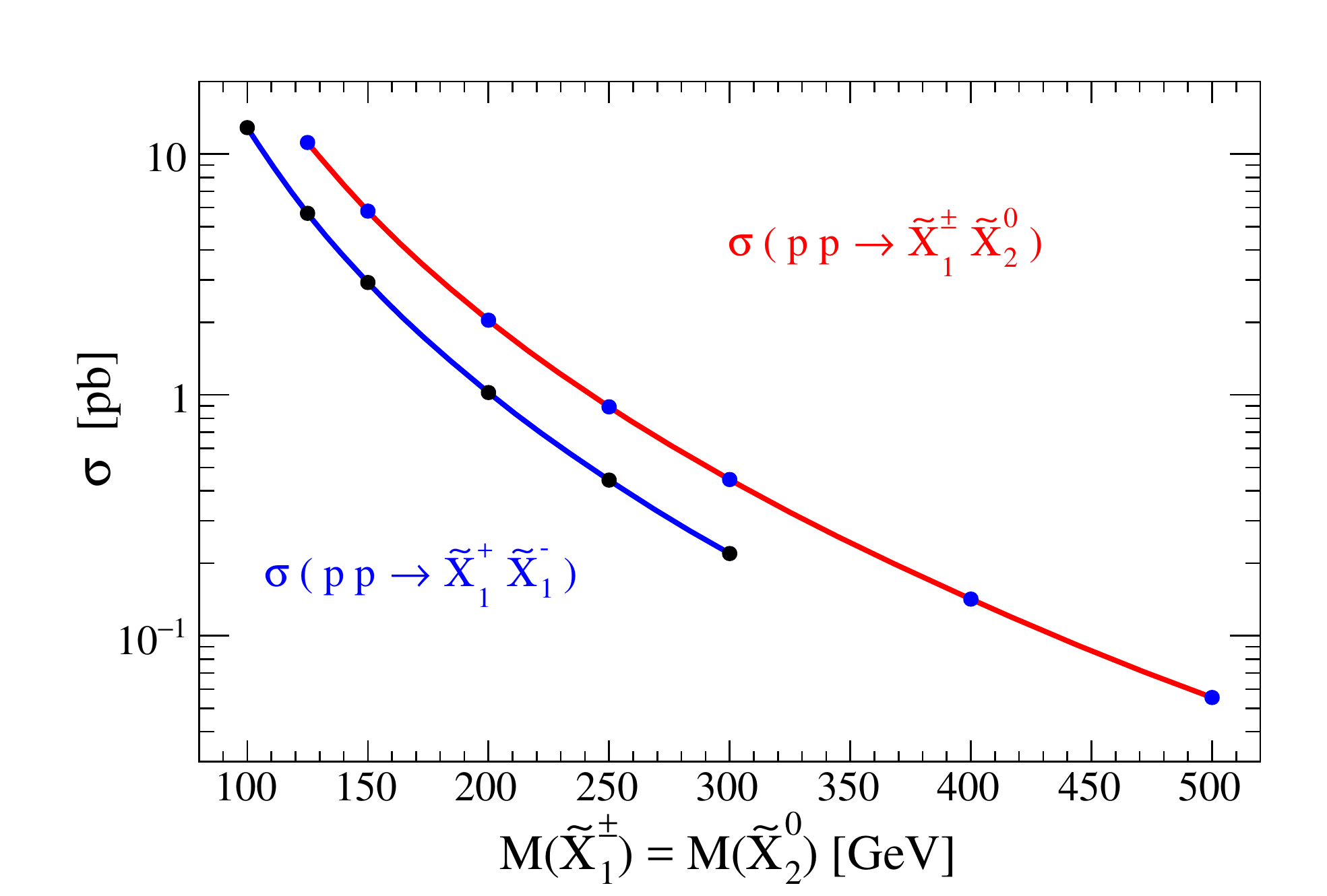}
\par\end{centering}

}\subfloat[\label{fig:aa}]{\noindent \begin{centering}
\includegraphics[height=3.2cm]{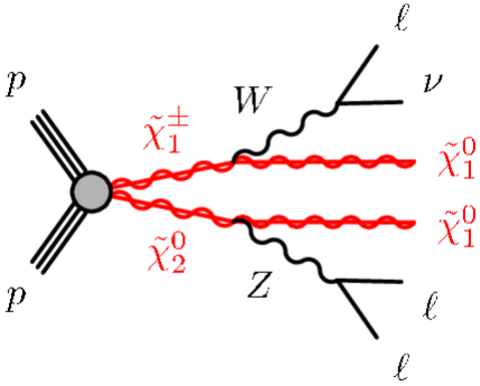}
\par\end{centering}

}\subfloat[\label{fig:bb}]{\noindent \begin{centering}
\includegraphics[height=3.2cm]{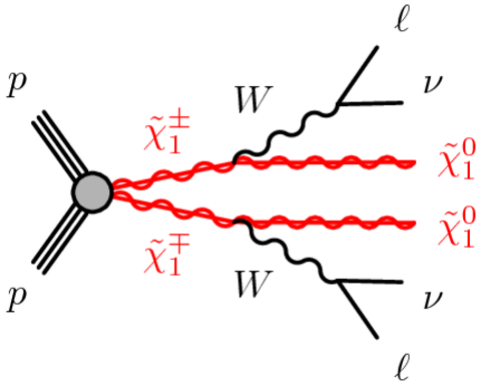}
\par\end{centering}

}\caption{Estimated NLL cross sections for pure wino chargino pair production
(blue curve) and chargino-neutralino associated production (red curve)
at $\sqrt{s}=14$ TeV (a). Feynman diagrams for electroweakino
productions in final states with missing transverse momentum and three charged
leptons (b) or two charged leptons (c). }
\end{figure}

The cross sections for pure wino chargino pair production and chargino-neutralino
associated production at $\sqrt{s}=13$ TeV at NLL can be found elsewhere
\cite{Fuks:2012qx,Fuks:2013vua} with relative uncertainties in the
range $4.5\%\lesssim\Delta\sigma\lesssim9\%$ for the masses investigated.
We estimate the NLL cross sections at $\sqrt{s}=14$
TeV, evaluating the NLO cross sections for the wino-like
electroweakino pair production at 13 and 14 TeV with Madgraph and assuming the
same NLL/NLO $k$-factors for the corrections. The resulting NLL cross
sections are shown in Figure \ref{fig:cc} and used as inputs for
the analysis of the simplified supersymmetric topologies. This procedure
provides small corrections ($\lesssim5\%$) from the $k$-factors and
is a check for the matched Madgraph cross sections and their potential dependences
on the cutoff scales chosen.

The focus of this work is on the leptonic decay channels of off-shell
$W$ and $Z$ bosons, as depicted in the Feynman diagrams in Figure
\ref{fig:aa} and Figure \ref{fig:bb}. Leptonic final states from charmonium
and bottomonium are expected to be negligible in the phase space probed.

Focusing on electrons and muons as visible decay products provides
several advantages. Firstly, the signal-to-background ratio 
increases progressively with lepton multiplicity
in the final state. Secondly, the channels result in clean final states
with high efficiencies for the lepton reconstruction: in this
study one assumes the minimum value for reconstructed
lepton $p_{T}$ of 10 GeV. Recent work by the CMS
collaboration has demonstrated improvements in the efficiency of
identification of \textit{soft} isolated electrons and muons (down to
$\sim$3-4 GeV) \cite{CMS:2016zvj}.

Moreover, for our purposes all leptons are identifiable as reconstructed
objects produced via sparticle decays and hence they are assigned
to the V-system, while all the jets can be assigned to the ISR-system
with no ambiguity. This allows us to avoid focusing solely on the high
ISR regime in order to exploit the compressed RJR strategy. A minimal
value of the transverse momentum of the ISR-system, in concert with
${\not\mathrel{E}}_{T}$, can elicit an increase in the
transverse momenta of the decay products of the SUSY system. In this way we can leverage
the RJR technique for compressed scenarios without requiring a
restrictive event selection based on a huge value of the ISR transverse
momentum.

\subsection{Chargino-neutralino associated production in final states with
three leptons and missing transverse momentum}

The signal samples are the simplified topologies as in Figure \ref{fig:aa}
generated within the mass ranges 125 GeV $\leq (M_{\tilde{\chi}_{1}^{\pm}}=M_{\tilde{\chi}_{2}^{0}})\leq$
500 GeV, with five mass splittings $\Delta M=M_{\tilde{P}}-M_{\tilde{\chi}_{1}^{0}}=15,$
25, 35, 50 and 75 GeV. Event-by-event a basis of RJR variables is
extracted and analyzed to probe compressed spectra for a projection
of $\int\mathcal{L}\,\mathrm{dt}=300\:\mathrm{fb}^{-1}$. To the previous variables,
additional transverse observables for this study include:
\begin{itemize}
\item $M_{T}^{\mathrm{V}}$: transverse mass of the V-system.
\item $M_{l^{+}l^{-}}$: transverse mass of the two same flavor opposite
sign leptons in final state where the third lepton has different flavor
($M_{T,\:e^{+}e^{-}}$ when the third lepton is a muon and $M_{T,\:\mu^{+}\mu^{-}}$
when the third lepton is an electron).
\item $\Delta\phi_{\mathrm{CM},\mathrm{I}}$: opening angle between the
CM system and the I-system.
\end{itemize}
Three leptons (electrons and muons) are required in the final state
with $p_{T}>10$ GeV, while at least one jet with $p_{T}>20$ GeV
is associated to the ISR-system. A minimal value for the transverse
missing momentum is the last preselection requirement: ${\not\mathrel{E}}_{T}>50$
GeV. 
\begin{figure}[t]
\noindent \begin{centering}
\subfloat[\label{fig:RISRCharNeutPre-1}]{\noindent \begin{centering}
\includegraphics[height=5cm]{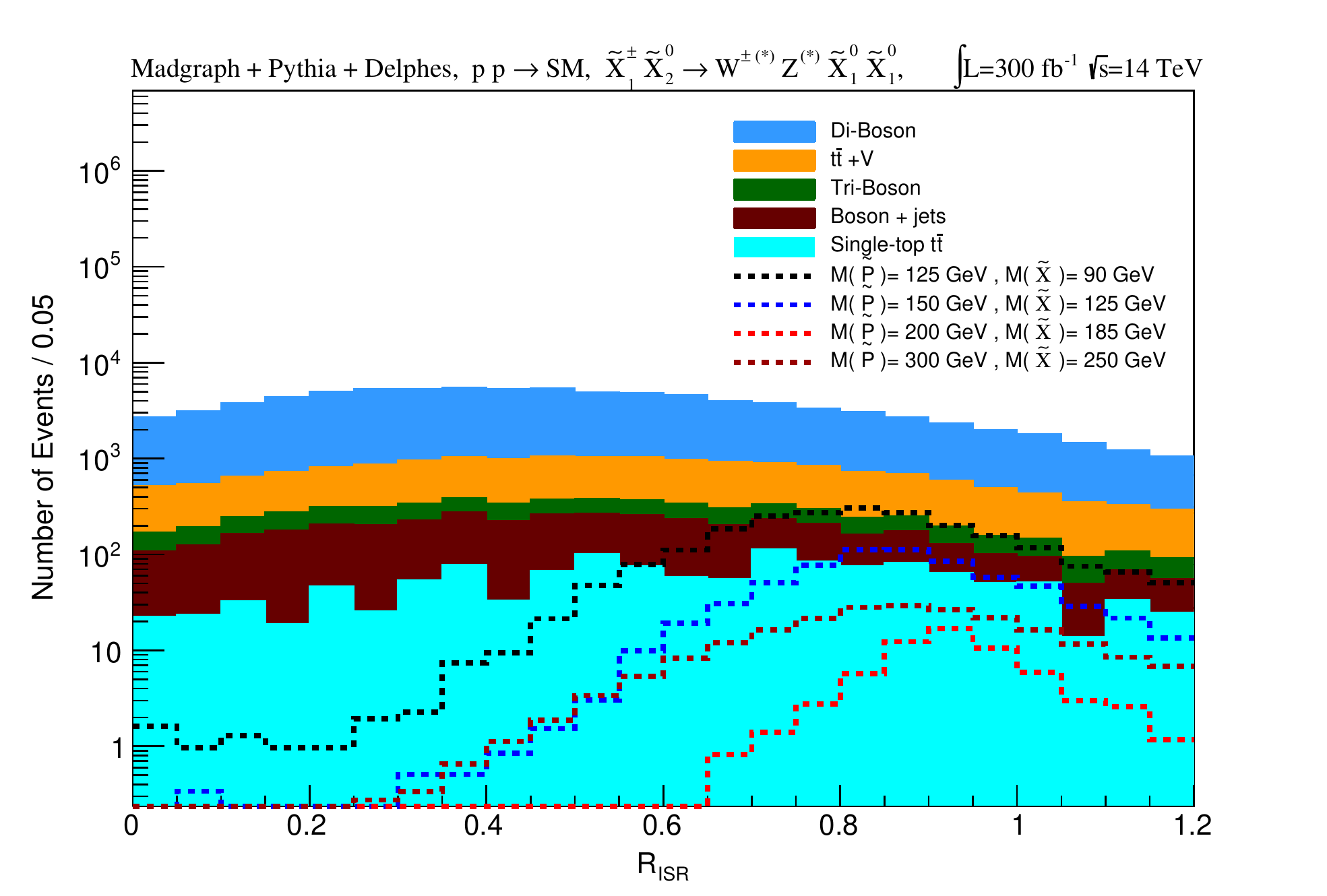}
\par\end{centering}

}\subfloat[\label{fig:PTISRCharNeutpre-1}]{\includegraphics[height=5cm]{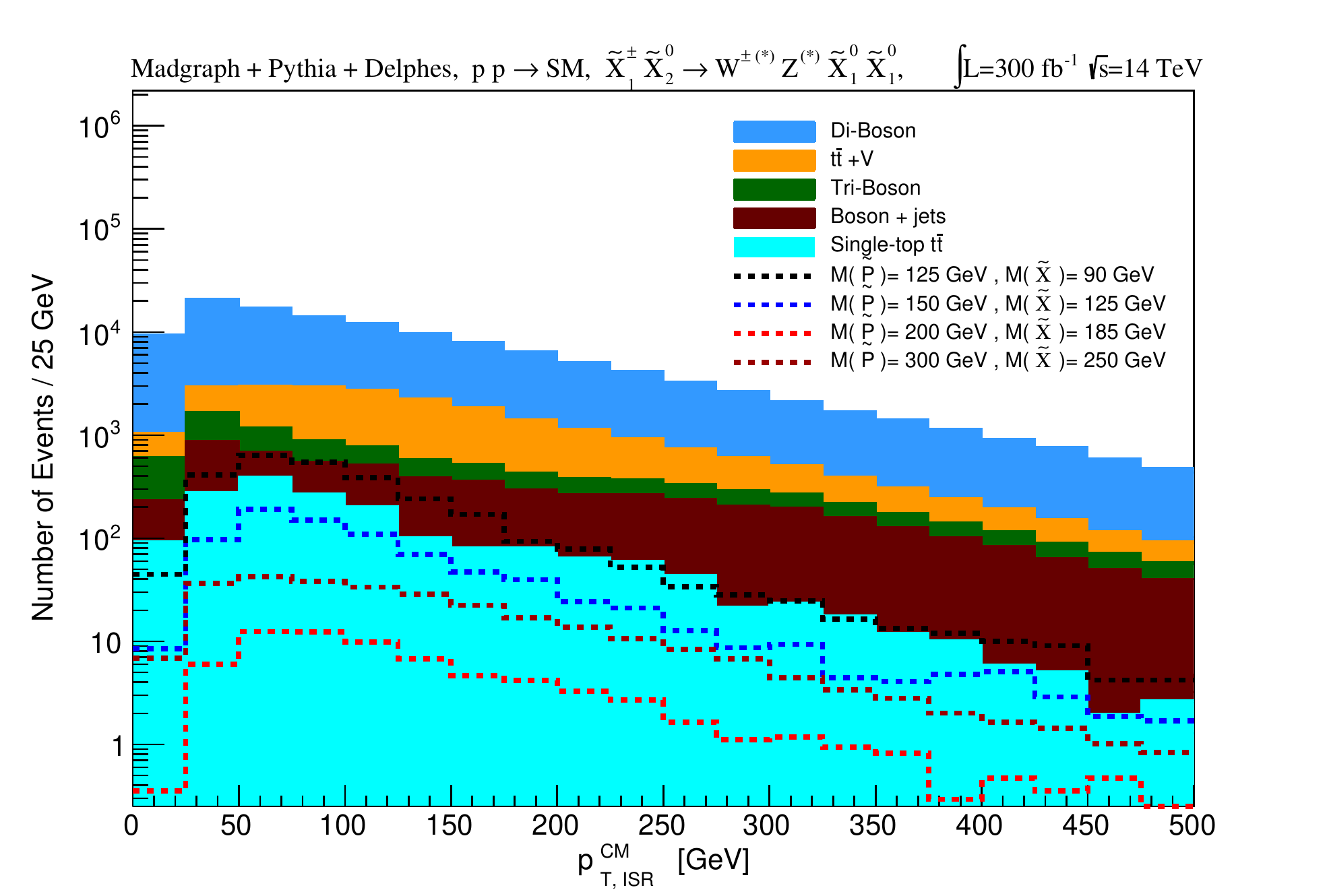}}
\par\end{centering}

\noindent \centering{}\caption{Distributions of $R_{\mathrm{ISR}}=\frac{\left|\vec{p}_{\mathrm{I},T}^{\mathrm{CM}}\cdot\hat{p}_{\mathrm{ISR},T}^{\mathrm{CM}}\right|}{p_{\mathrm{ISR},T}^{\mathrm{CM}}}$
(a) and $p_{\mathrm{ISR},T}^{\mathrm{CM}}$
(b) for events that have satisfied the
preselection criteria. All the contributions are scaled with an integrated
luminosity of 300 $\mathrm{fb}^{-1}$ at $\sqrt{s}$=14 TeV. \label{fig:Distribution-CharNeut-1}}
\end{figure}

Figure \ref{fig:Distribution-CharNeut-1} shows the distributions
of $R_{\mathrm{ISR}}$ and $p_{\mathrm{ISR},T}^{\mathrm{CM}}$ after preselection 
criteria are imposed. All of
the relevant Standard Model backgrounds are stacked together and categorized
into five groups. The main contributions are $WZ$ boson associated
production and $t\bar{t}$ processes with an additional vector
boson. The overlaid dashed curves refer to four chargino-neutralino
production samples with different masses and mass splittings. 

The observable $R_{\mathrm{ISR}}$ provides a remarkable signal-to-background
discrimination in the absence of more stringent selection criteria
as shown in Figure \ref{fig:RISRCharNeutPre-1}. The assignment of
the different objects in the compressed tree is performed with no ambiguity,
and it is not necessary to focus on the high ISR regime in order to
improve the observable resolution for the signal samples. Notice that
$R_{\mathrm{ISR}}$ can assume larger values than unity when some objects
are forced in the $V$-system. The observable is expected to be peaked
for values beyond $M_{\tilde{\chi}}/M_{\tilde{P}}$ due to the
additional contribution to ${\not\mathrel{E}}_{T}$ coming from one
or more neutrinos. Values larger than the mass ratio will be considered  for the definition of the $R_{\mathrm{ISR}}$
requirements together with $R_{\mathrm{ISR}}<1$. 

Figure \ref{fig:PTISRCharNeutpre-1} shows the distribution of $p_{\mathrm{ISR},T}^{\mathrm{CM}}$.
The scales for the signal and background samples are similar
and the variable has limited impact. In the absence of other requirements
the slope for the signal is paradoxically more severe arising from the
background events with non-radiative jets, forced into the ISR-system.
A minimal requirement on $p_{\mathrm{ISR},T}^{\mathrm{CM}}$ is essential
to exploit the RJR technique with multi-lepton final states. The
requirement applied to this variable, being the only large scale observable
in this study together with ${\not\mathrel{E}}_{T}$, will be moderately
tighter for large mass splittings, when the criterion on $R_{\mathrm{ISR}}$
is relaxed.

It is interesting to note that the number of events passing the preselection criteria
is smaller for the signal sample with $M_{\tilde{\chi}_{1}^{\pm}}=M_{\tilde{\chi}_{2}^{0}}$=200
GeV and $\Delta M$ =15 GeV compared to the sample with  $M_{\tilde{\chi}_{1}^{\pm}}=M_{\tilde{\chi}_{2}^{0}}$=300
GeV and $\Delta M$ =50 GeV. There has been a conservative minimal choice of 
transverse momentum for electrons and muons of 10 GeV and when the
mass splitting approaches a much more compressed regime, the kinematics are such that
one of the three leptons is less likely to satisfy this transverse momentum
constraint. In order to probe the extreme compressed regime ($\Delta M<$ 15 GeV)
a parametrization of the efficiency in the reconstruction of soft
electron and muons with transverse momenta ($p_{T}\lesssim10$ GeV)
must be implemented. This is considered beyond the scope of this paper due to the difficulty 
of getting these details correct outside of an experimental collaboration.

\begin{figure}[t]
\noindent \begin{centering}
\subfloat[\label{fig:RMTVDiBoson}]{\noindent \begin{centering}
\includegraphics[width=0.32\textwidth]{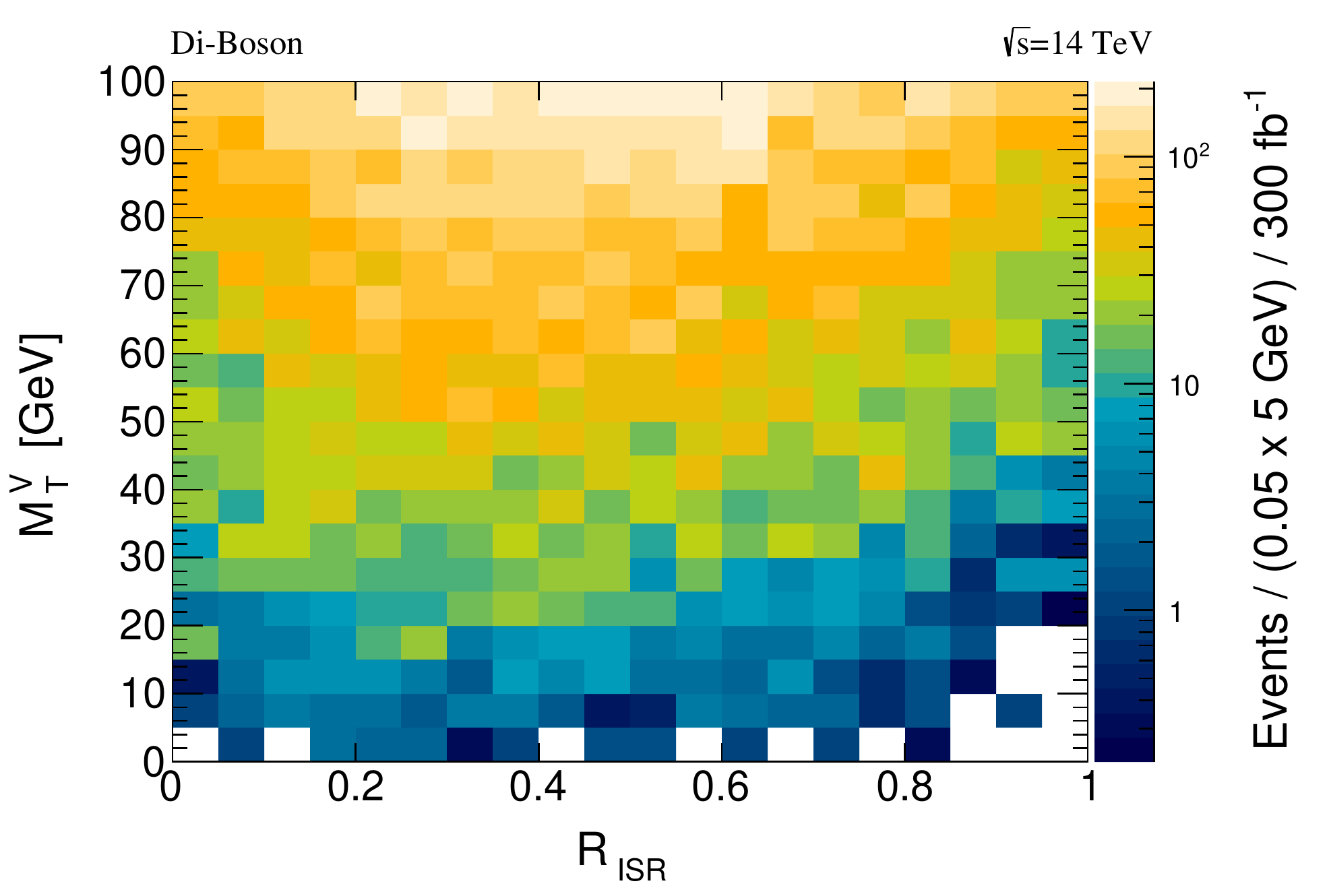}
\par\end{centering}

}\subfloat[\label{fig:RMTVD15}]{\noindent \begin{centering}
\includegraphics[width=0.32\textwidth]{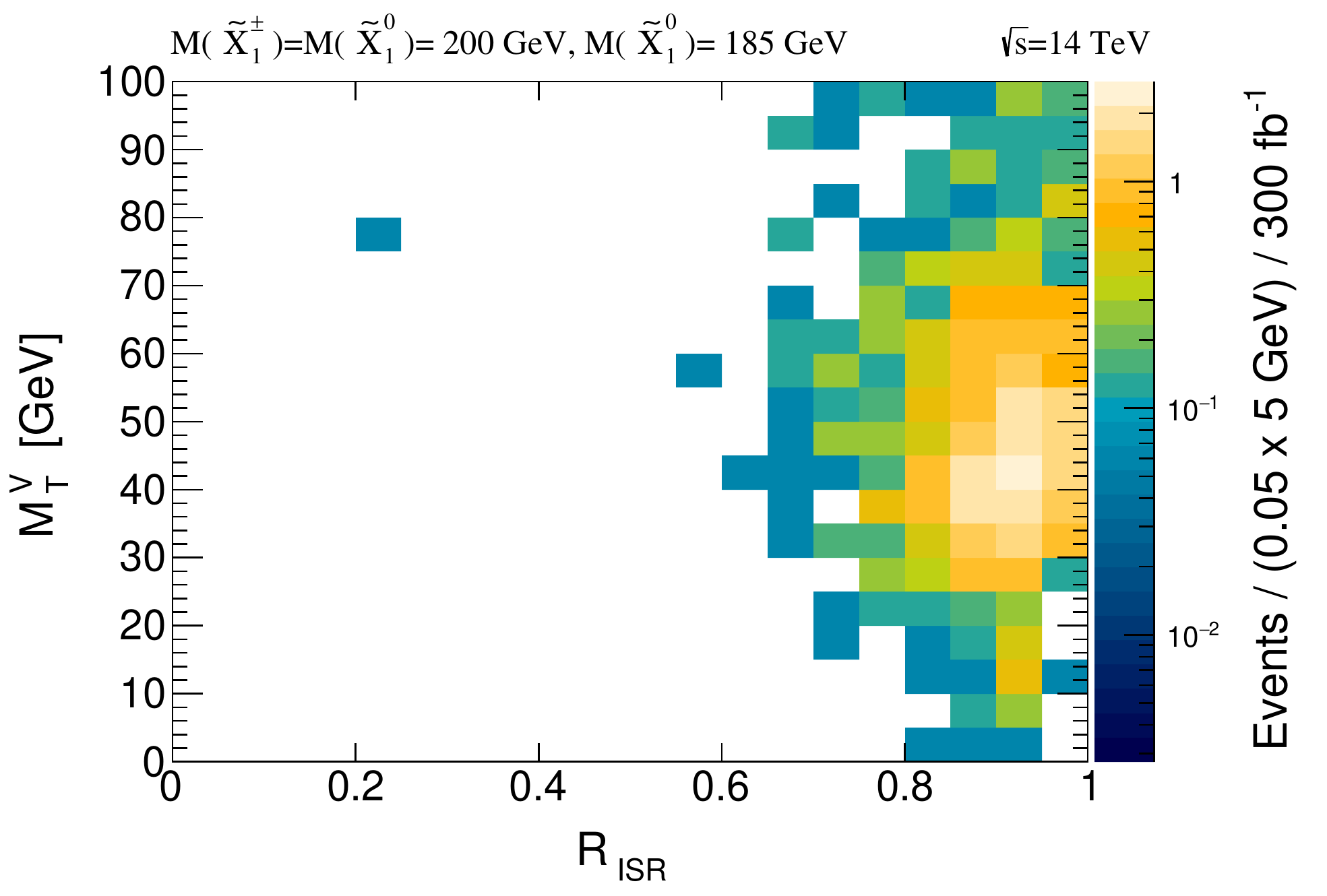}
\par\end{centering}

}\subfloat[\label{fig:RMTVD25}]{\noindent \begin{centering}
\includegraphics[width=0.32\textwidth]{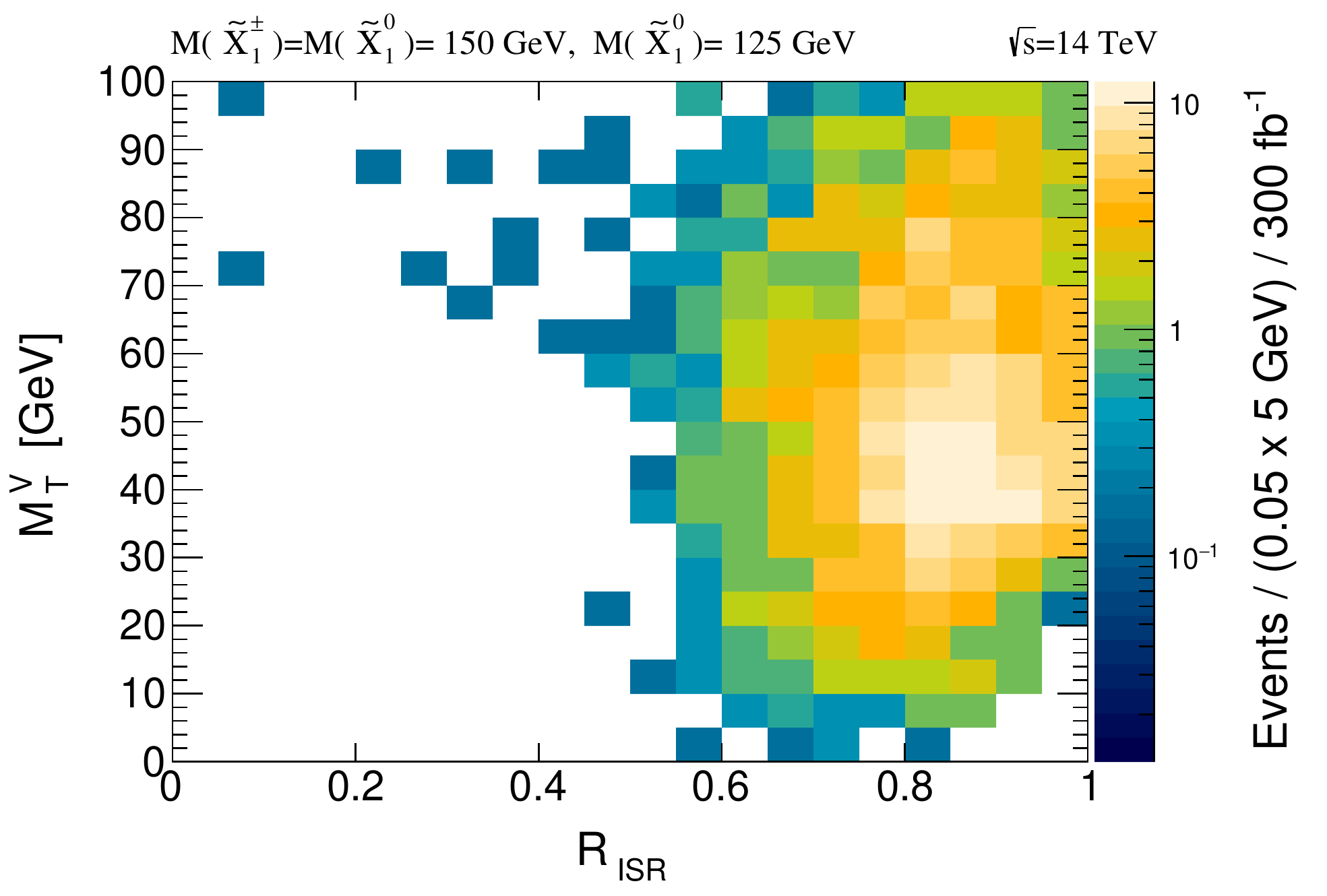}
\par\end{centering}

}
\par\end{centering}

\noindent \centering{}\caption{Distribution of the $M_{T}^{\mathrm{V}}$ as a function of $R_{\mathrm{ISR}}$
for the di-boson background (a) and the signal
samples $M_{\tilde{\chi}_{1}^{\pm}}=M_{\tilde{\chi}_{2}^{0}}$=200
GeV, $M_{\tilde{\chi}_{1}^{0}}$=185 GeV (b) and
$M_{\tilde{\chi}_{1}^{\pm}}=M_{\tilde{\chi}_{2}^{0}}$=150 GeV, $M_{\tilde{\chi}_{1}^{0}}$=125
GeV (c). One demands preselection criteria, $N_{b-\mathrm{jet}}^{\mathrm{ISR}}=0$,
$p_{\mathrm{ISR},T}^{\mathrm{CM}}>50\:\mathrm{GeV}$ and $M_{T}^{\mathrm{V}}<100$
GeV.\label{fig:RMTVBG}}
\end{figure}

Figure \ref{fig:RMTVBG} shows the two dimensional distributions of
$M_{T}^{\mathrm{V}}$ as a function of $R_{\mathrm{ISR}}$ for the
main Standard Model background and two representative signal samples for events passing
the preselection criteria, and after applying a veto for jets tagged
as being initiated by a $b$-quark ($N_{b-jet}^{\mathrm{ISR}}=0$). The final state signal events
populate low values of $M_{T}^{\mathrm{V}}$ with a complementarity
with high values of $R_{\mathrm{ISR}}$. Vice versa, for the di-boson
background, simultaneous low values of $M_{T}^{\mathrm{V}}$
and $R_{\mathrm{ISR}}$ close to one are disfavored as shown in
Figure \ref{fig:RMTVDiBoson}.

\begin{figure}[t]
\noindent \begin{centering}
\subfloat[\label{fig:dphiISRINC} ]{\noindent \begin{centering}
\includegraphics[width=0.32\textwidth]{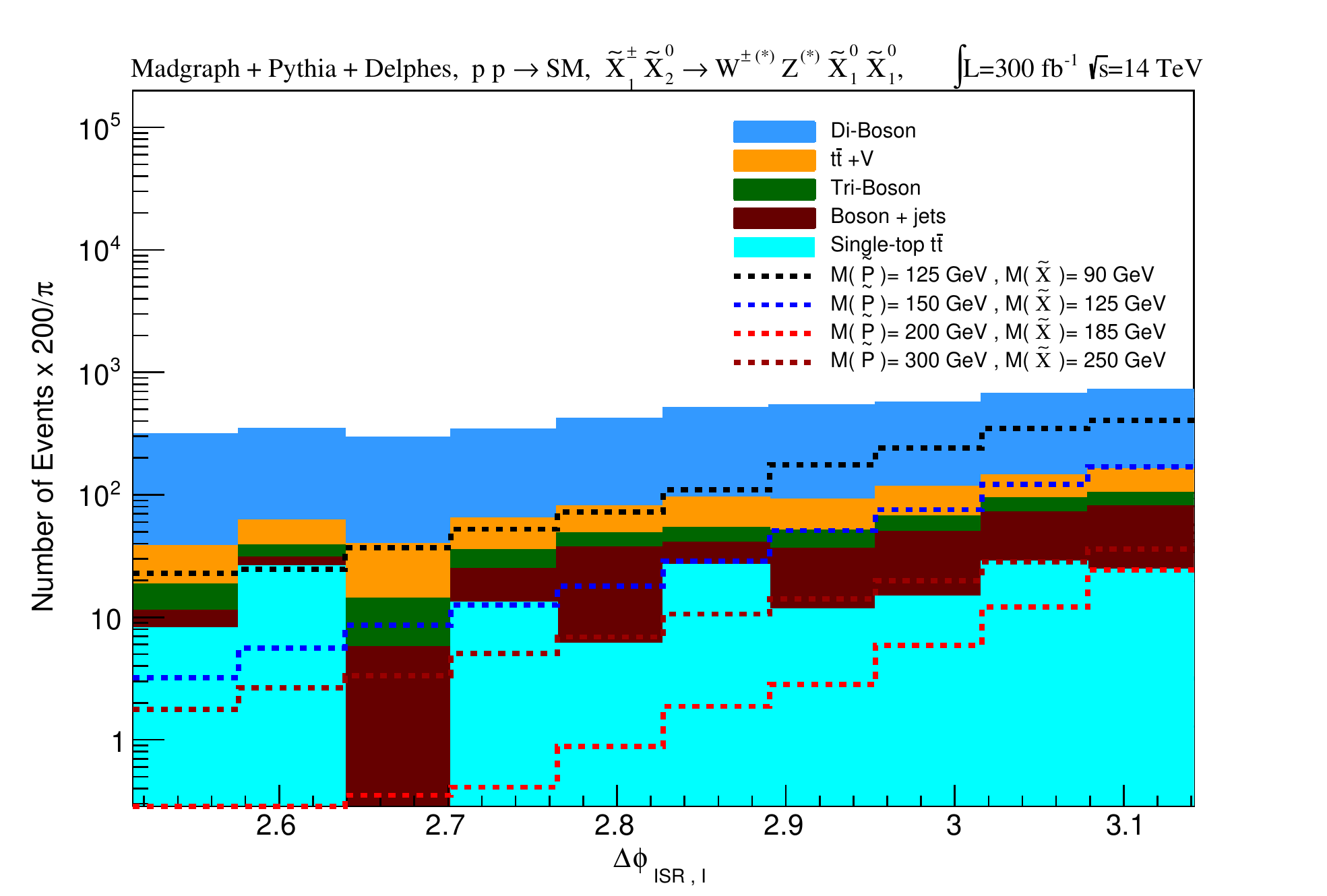}
\par\end{centering}

}\subfloat[\label{fig:2DdphiISRRLowMTV-4} ]{\noindent \begin{centering}
\includegraphics[width=0.32\textwidth]{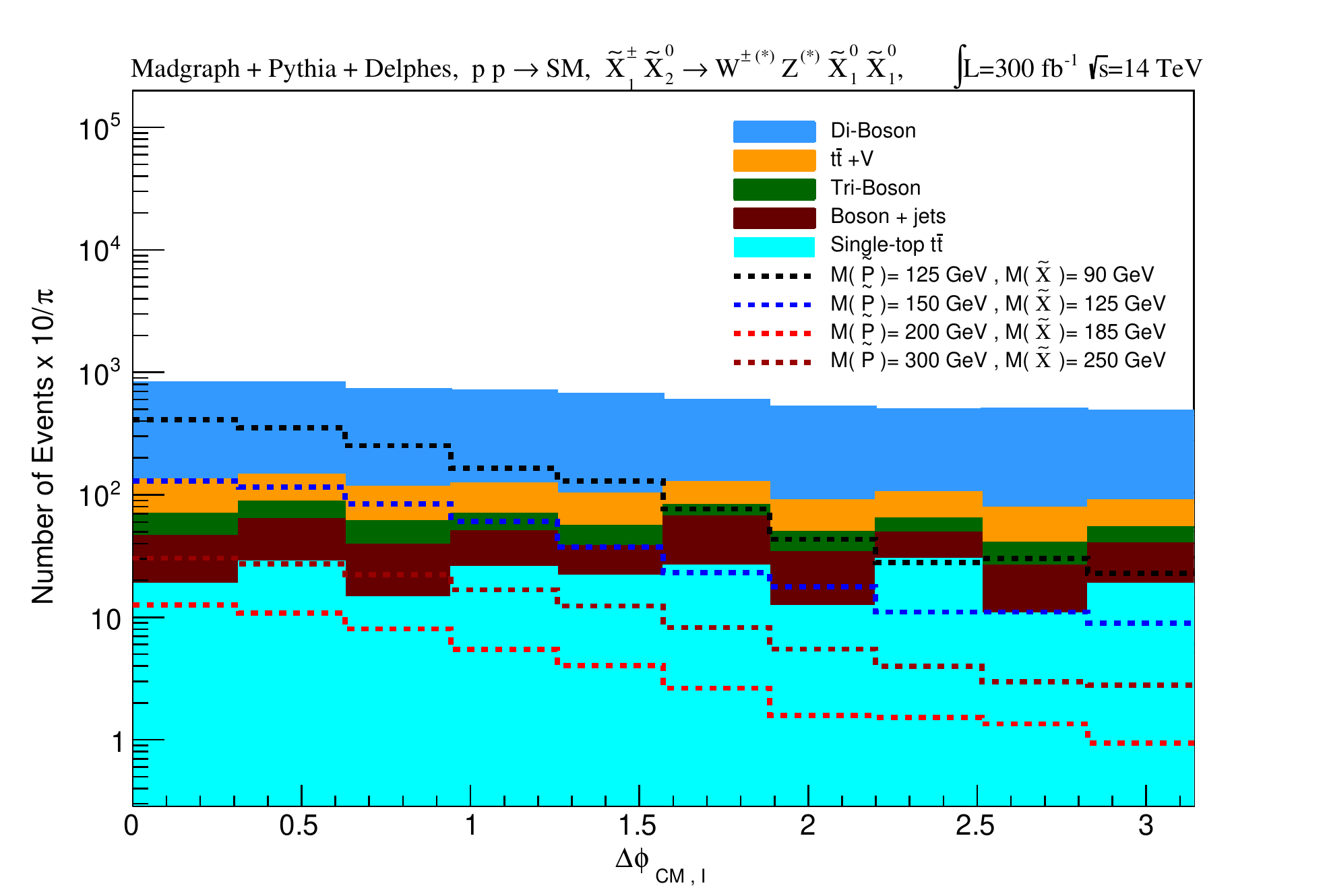}
\par\end{centering}

}\subfloat[\label{fig:2DdphiISRRLowMTV-4-2} ]{\noindent \begin{centering}
\includegraphics[width=0.32\textwidth]{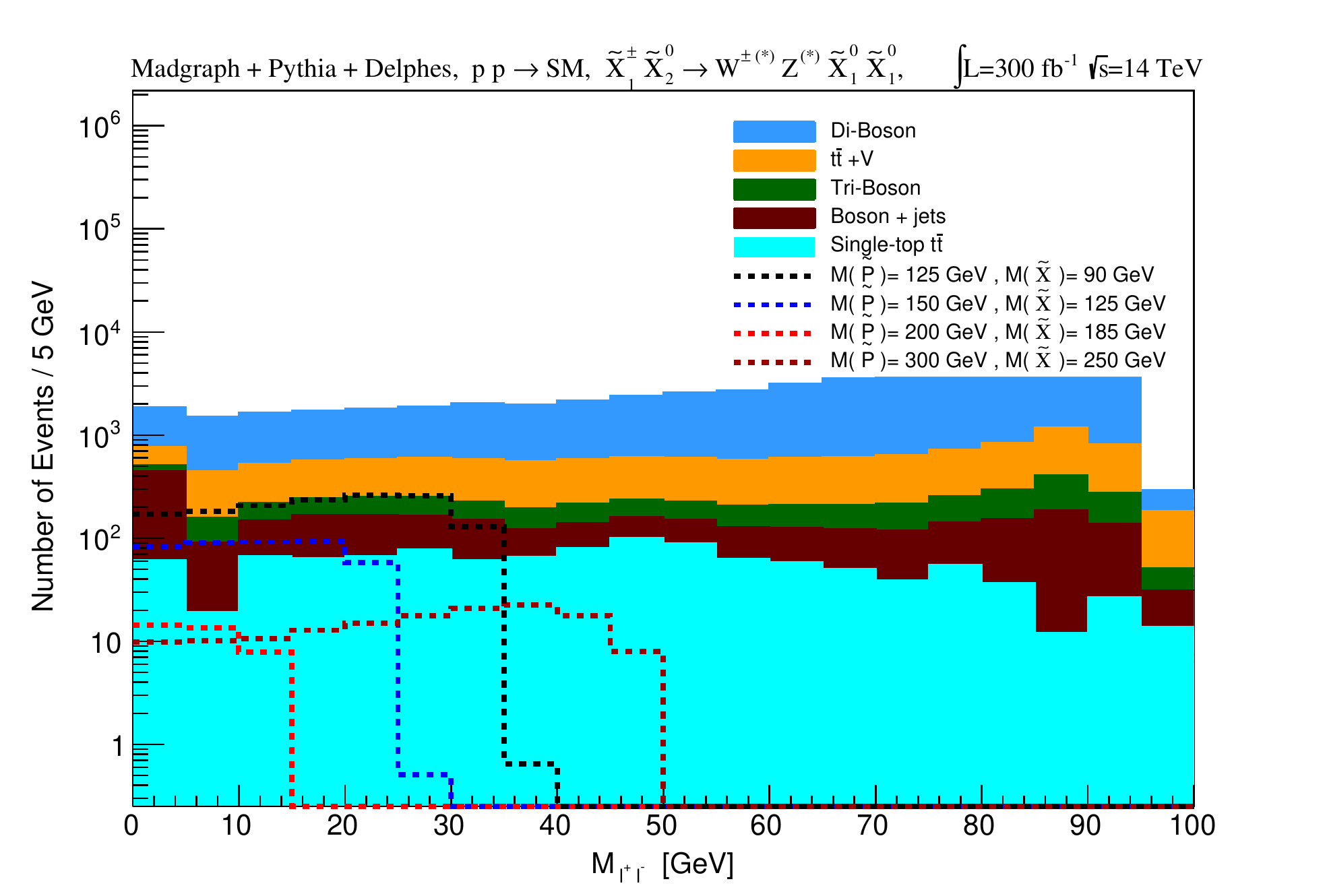}
\par\end{centering}

}
\par\end{centering}

\noindent \centering{}\caption{Distributions of $\Delta\phi_{\mathrm{ISR},\mathrm{I}}$ (a),
$\Delta\phi_{\mathrm{CM},\mathrm{I}}$(b)
and $M_{l^{+}l^{-}}$ (c) for the events
satisfying the preselection criteria, $N_{b-\mathrm{jet}}^{\mathrm{ISR}}=0$,
$p_{\mathrm{ISR},T}^{\mathrm{CM}}>50\:\mathrm{GeV}$, $M_{T}^{\mathrm{V}}<100$
GeV and $R_{\mathrm{ISR}}>0.6$. \label{fig:NCdphidphiMLL}}
\end{figure}

Using the two RJR observables in concert provides an increasingly powerful
discrimination the smaller the absolute and relative mass splitting of
the signal sample. In the low $M_{T}^{\mathrm{V}}$ regime ($M_{T}^{\mathrm{V}}<100$~GeV), and for values of the ratio close to unity ($R_{\mathrm{ISR}}>0.6$)
additional handles to decrease the SM background yields are provided
by the compressed-transverse RJR angles and $M_{l^{+}l^{-}}$. Figure
\ref{fig:NCdphidphiMLL} shows the distribution of $\Delta\phi_{\mathrm{ISR},\mathrm{I}}$
(a), $\Delta\phi_{\mathrm{CM},\mathrm{I}}$ (b) and $M_{l^{+}l^{-}}$
for final state events with only two of the three leptons with same
flavor (c). For the signal events, the transverse mass of the two
leptons has a clean end-point at the expected $\Delta M$.

Selection criteria applied on the compressed RJR observables, as shown in Table \ref{tab:A-SignRegNC}, 
can be used to define signal regions for probing chargino-neutralino associated 
pair production in final states with three leptons and missing transverse momentum. 
One or more additional jets is assumed to be radiated
from the initial state and a minimal requirement on $p_{\mathrm{ISR},T}^{\mathrm{CM}}$
(and ${\not\mathrel{E}}_{T}$) allows us to focus on the final states
of interest and probe the compressed spectra. The signal regions target
five particular mass splittings. A special treatment is assumed
for the selection criteria applied to $R_{\mathrm{ISR}}$, since this observable is 
related to the mass ratio $M_{\tilde{\chi}}/M_{\tilde{P}}$
rather then the absolute value of the mass splitting.
\begin{table}[tp]
\begin{tabular}{|>{\centering}m{0.27\textwidth}|>{\centering}p{0.11\textwidth}|>{\centering}p{0.11\textwidth}|>{\centering}p{0.11\textwidth}|>{\centering}p{0.11\textwidth}|>{\centering}p{0.11\textwidth}|}
\cline{2-6} 
\multicolumn{1}{>{\centering}m{0.27\textwidth}|}{\foreignlanguage{british}{}} & \multicolumn{5}{c|}{{\small{}$\begin{array}{c}
\\
\\
\end{array}$}\textbf{Mass Splitting {[}GeV{]}}{\small{}$\begin{array}{c}
\\
\\
\end{array}$}}\tabularnewline
\hline 
{\small{}$\begin{array}{c}
\\
\\
\end{array}$}\textbf{Variable}{\small{}$\begin{array}{c}
\\
\\
\end{array}$} & {\small{}$\Delta M=15$} & {\small{}$\Delta M=25$} & {\small{}$\Delta M=35$} & {\small{}$\Delta M=50$} & {\small{}$\Delta M=75$}\tabularnewline
\hline 
\multicolumn{6}{c}{
{\small{}\vspace{-0.4cm}
}\selectlanguage{english}%
}\tabularnewline
\hline 
\foreignlanguage{british}{{\small{}Object multiplicity}} & \multicolumn{5}{c|}{{\small{}3 Leptons ($e$ and $\mu$) with $p_{T}^{lep}>10$ GeV, }}\tabularnewline
{\small{}selection criteria} & \multicolumn{5}{c|}{{\small{}At least one jet, $p_{T}^{\:jet}>20$ GeV, $N_{b-jet}^{\mathrm{ISR}}=0$}}\tabularnewline
\hline 
{\small{}$p_{\mathrm{ISR},T}^{\mathrm{CM}}$ (${\not\mathrel{E}}_{T}$)
$>$ {[}GeV{]}} & \multicolumn{3}{c|}{{\small{}$50$}} & {\small{}$70$} & {\small{}$120$}\tabularnewline
\hline 
{\small{}$N_{jet}^{\mathrm{ISR}}<$} & {\small{}$3$} & \multicolumn{2}{c|}{{\small{}$4$}} & \multicolumn{2}{c|}{{\small{}$3$}}\tabularnewline
\hline 
\multicolumn{1}{|>{\centering}m{0.27\textwidth}||}{{\small{}$M_{T}^{\mathrm{V}}<$, for 3 SFL {[}GeV{]}}} & {\small{}$40$} & {\small{}$50$} & {\small{}$65$} & {\small{}$75$} & \multicolumn{1}{>{\centering}p{0.1\textwidth}|}{{\small{}$90$}}\tabularnewline
\hline 
\multicolumn{1}{|>{\centering}m{0.27\textwidth}||}{{\small{}$M_{l^{+}l^{-}}<$, for 2 SFL {[}GeV{]} ($M_{T}^{\mathrm{V}}<100$
GeV)}} & {\small{}$15$} & {\small{}$25$} & {\small{}$35$} & {\small{}$50$} & {\small{}$75$}\tabularnewline
\hline 
{\small{}$\Delta\phi_{\mathrm{CM},\mathrm{I}}<$} & \multicolumn{3}{c|}{{\small{}$1$}} & {\small{}$0.7$} & {\small{}$0.5$}\tabularnewline
\hline 
{\small{}$\Delta\phi_{\mathrm{ISR},\mathrm{I}}>$} & \multicolumn{5}{c|}{{\small{}$3$}}\tabularnewline
\hline 
{\small{}$R_{\mathrm{ISR}}>$} & {\small{}$0.85,\;0.9$} & {\small{}$\begin{array}{c}
0.8,\;0.85\\
0.9
\end{array}$} & \foreignlanguage{british}{{\small{}$0.8,\;0.85$}} & {\small{}$\begin{array}{c}
0.7,\;0.8\\
0.85
\end{array}$} & {\small{}$\begin{array}{c}
0.65,\;0.7\\
0.75
\end{array}$}\tabularnewline
\hline 
\end{tabular}

\caption{A loosely optimized set of selection criteria for signal regions
in the analysis of chargino neutralino production in final states
with three leptons and missing energy. \label{tab:A-SignRegNC}}
\end{table}
\begin{figure}[t]
\noindent \begin{centering}
\subfloat[]{\noindent \begin{centering}
\includegraphics[height=5cm]{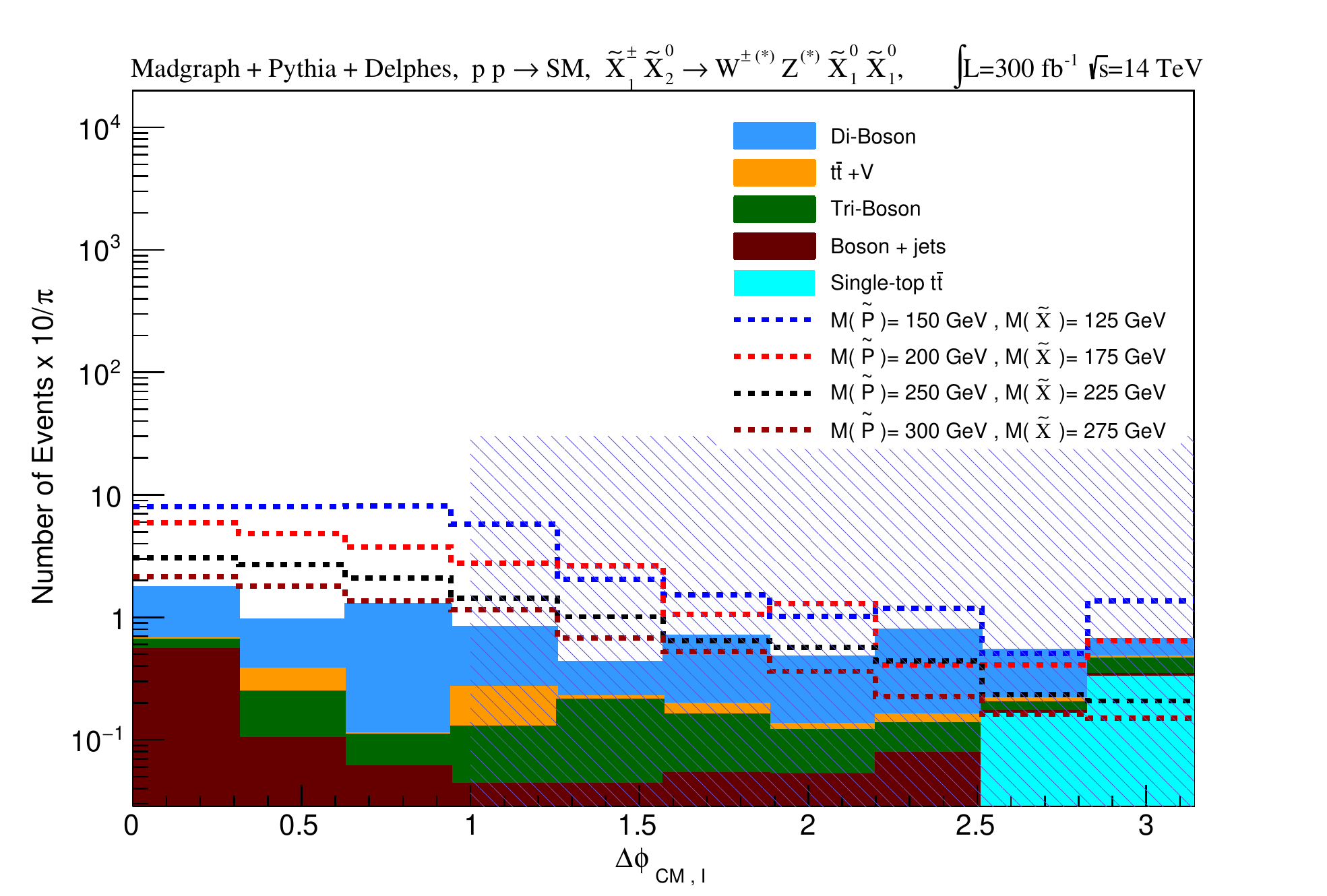}
\par\end{centering}

}\subfloat[]{\includegraphics[height=5cm]{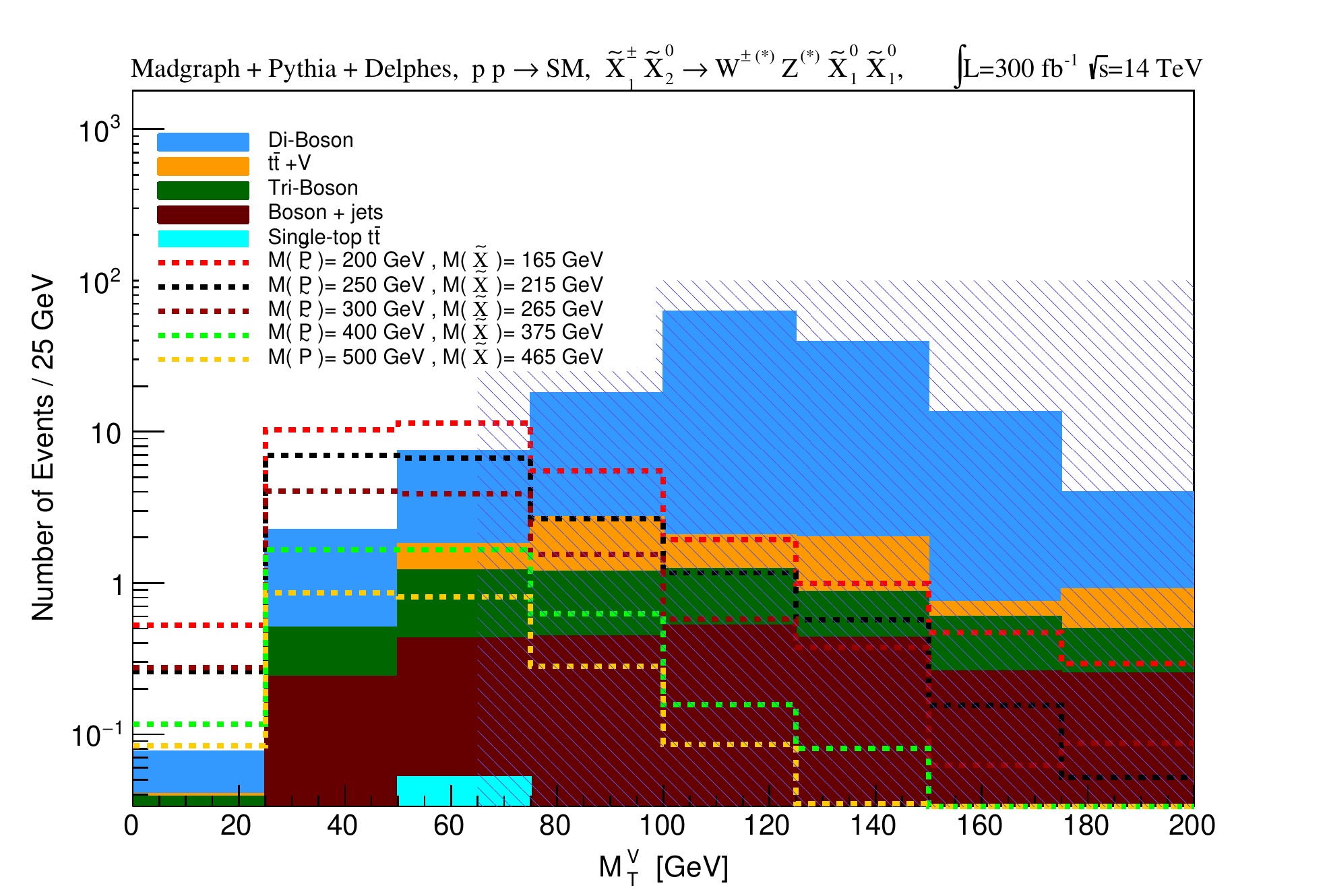}}
\par\end{centering}

\noindent \centering{}\caption{The distributions of the observables $\Delta\phi_{\mathrm{CM},\mathrm{I}}$
and $M_{T}^{\mathrm{V}}$ for signal and Standard Model background
events passing the N-1 selection criteria in one of the signal regions
of Table \ref{tab:A-SignRegNC} with $R_{\mathrm{ISR}}$>0.85.\label{fig:DistributionN-1NC} }
\end{figure}
Tighter selection criteria are used for the only 
large scale variables ( $p_{\mathrm{ISR},T}^{\mathrm{CM}}$
and ${\not\mathrel{E}}_{T}$), the jet multiplicity and $\Delta\phi_{\mathrm{CM},\mathrm{I}}$
when for larger mass differences ($\Delta M=50$, 75 GeV) the $R_{\mathrm{ISR}}$
requirement is relaxed. Low values are required for $M_{T}^{\mathrm{V}}$,
progressively more stringent to the decrease of $\Delta M$, while
for $M_{l^{+}l^{-}}$, one requires a maximum defined exactly by the
mass splitting itself. In final states with three electrons or three
muons only the $M_{T}^{\mathrm{V}}$ requirement is applied, while
for events with two same and one different flavor leptons the selection
on $M_{l^{+}l^{-}}$ is required together with $M_{T}^{\mathrm{V}}<$100
GeV. The selection criteria applied to the observable $R_{\mathrm{ISR}}$
are progressively more stringent the closer the mass ratio to unity and
the values are separated by 0.05, which provides a moderate optimization.
Figure \ref{fig:DistributionN-1NC} shows the distributions of $\Delta\phi_{\mathrm{CM},\mathrm{I}}$
and $M_{T}^{\mathrm{V}}$ applying respectively the N-1 requirements
in column 2 and 3 of Table \ref{tab:A-SignRegNC}.

\begin{figure}[t]
\noindent \centering{}\includegraphics[height=8cm]{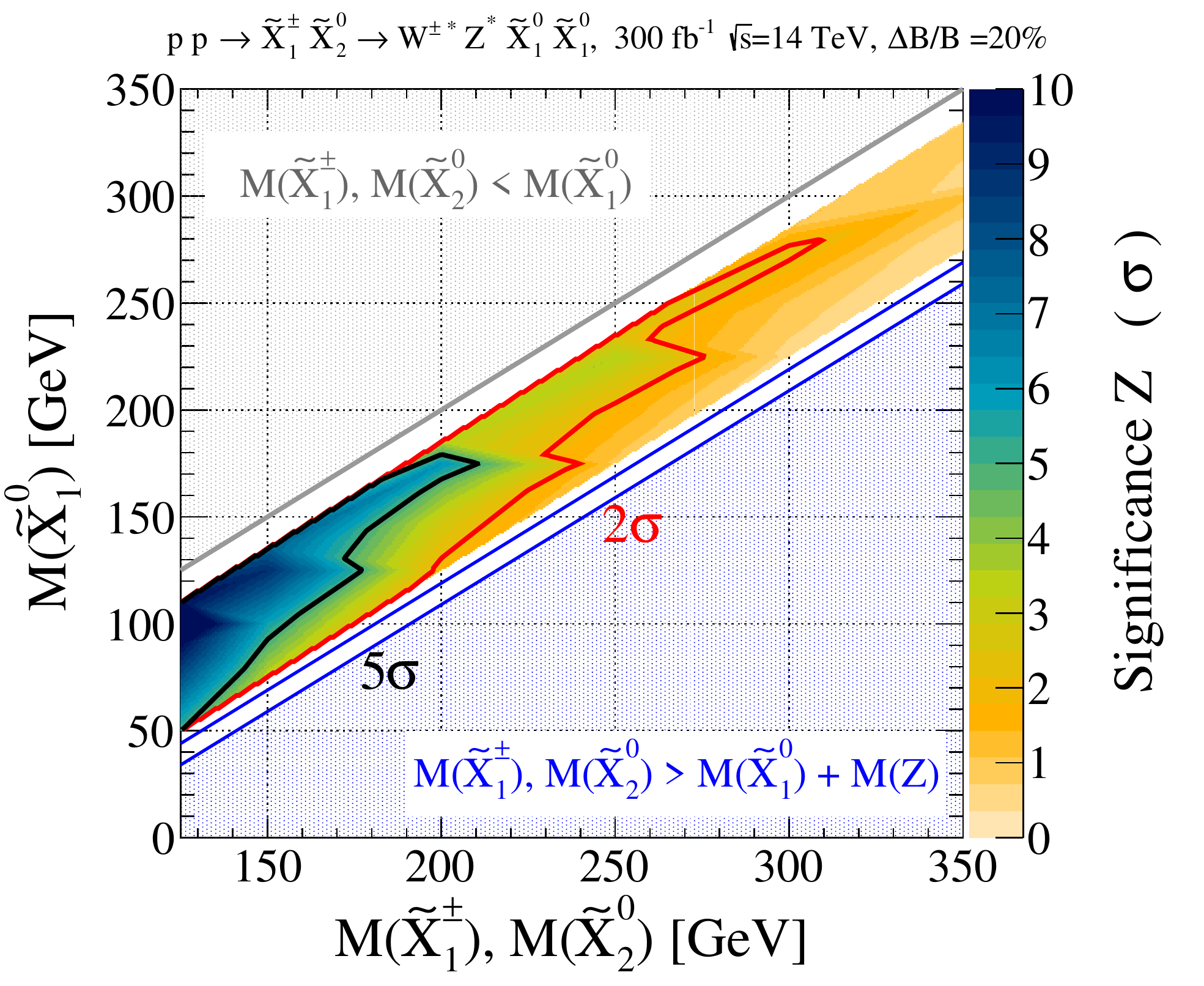}\caption{Projected exclusion and discovery reach for chargino-neutralino associated
production in the compressed region (15 GeV $\leq\Delta M\leq$ 75
GeV) at $\sqrt{s}=$ 14 TeV for an integrated luminosity of 300 $\mathrm{fb}^{-1}$.
\label{fig:ZValue_100fb_NC}}
\end{figure}

The signal regions expressed by the selection criteria of the RJR
observables defined in Table \ref{tab:A-SignRegNC} are applied to
calculate projected sensitivities for compressed spectra signal samples.
Figures \ref{fig:ZValue_100fb_NC} shows the value of $Z_{Bi}$
calculated assuming the metric \cite{Cousins:2008zz} at $\sqrt{s}$=14
TeV for an integrated luminosity of 300 $\mathrm{fb}^{-1}$. A systematic
uncertainty of 20\% is assumed for the whole signal grid with
the main contribution to the background arising from associated \textit{WZ} production.

The highest impact of the compressed RJR observables are for the samples
of mass splittings in the range 20-40 GeV, 
a challenging phase space for SUSY searches \cite{Aad:2014vma}.  
The signal yields in the
extreme compressed scenarios can benefit from an improvement in the
efficiencies of the detector in the reconstruction of low-momentum
leptons, which is outside the scope of this work. On the other hand,
the significances decrease for mass differences close to the $W$
pole mass due the difficulty to discriminate background events derived
from topologies with absolute and relative mass scales very close
to the signal ones. 

For an integrated luminosity of 300 $\mathrm{fb}^{-1}$, degenerate charginos
and neutralinos would be discovered for masses $M_{\tilde{\chi}_{1}^{\pm}}=M_{\tilde{\chi}_{2}^{0}}$>150
GeV for a large portion of the samples investigated and excluded up
to 300 GeV for the best scenarios. 

The value $\Delta M=$15 GeV must
not be considered as a threshold: the minimum mass difference achievable
with any technique is strongly related to the efficiencies for the
detector to reconstruct low-momentum leptons. For extremely compressed
scenarios, a similar analysis could be used to probe the same final
state topologies with only two low-momentum leptons reconstructed. 
Although the background would differ in that case, 
one could require two same-sign leptons to suppress the SM yield. 
Overall, one can improve the impact of the RJR technique by adopting
a strategy based not only on transverse observables, but exploiting 
a three-dimensional reconstruction as in the following study.

\subsection{Chargino pair production in final states with two leptons and missing
transverse momentum}

We now move on to consider a different and still more 
complicated compressed electroweakino investigation.
The signal samples are simulated proton-proton collisions at $\sqrt{s}=14$
TeV producing a pair of charginos with opposite electric charge and with $\tilde{\chi}_{1}^{\pm}\rightarrow W^{*\pm}(\rightarrow l^{\pm}\nu)\tilde{\chi}_{1}^{0}$. 
The focus is on final states with two leptons as illustrated in the simplified
topology in Figure \ref{fig:bb}. The samples are generated within
the mass ranges 100 GeV $\leq M_{\tilde{\chi}_{1}^{\pm}}\leq300$
GeV, with the five mass splittings $\Delta M=M_{\tilde{P}}-M_{\tilde{\chi}}=15,$
25, 35, 50 and 75 GeV. 

The lepton multiplicity of the final state determines the main contributions
of the Standard Model processes. The di-leptonic channels of a pair
of $W$ bosons, constitute the main processes resulting in a final state
with two opposite sign leptons and missing transverse momentum, in
the absence of hadronic jets. Searches for chargino pair production
in final state with two leptons are challenging for open mass spectra
due to the $W^{+}W^{-}$ irreducible background, while other contributions
are often negligible. In the compressed regime the difficulty is exacerbated
by the low momenta of invisible and visible objects and the subsequent
kinematics. Moreover, requiring a transverse momentum for the ISR-system 
introduces an additional complication for the analysis in
the compressed regime: Standard Model backgrounds other than $WW$ will contribute quite significantly.
\begin{figure}[t]
\noindent \centering{}\includegraphics[height=5cm]{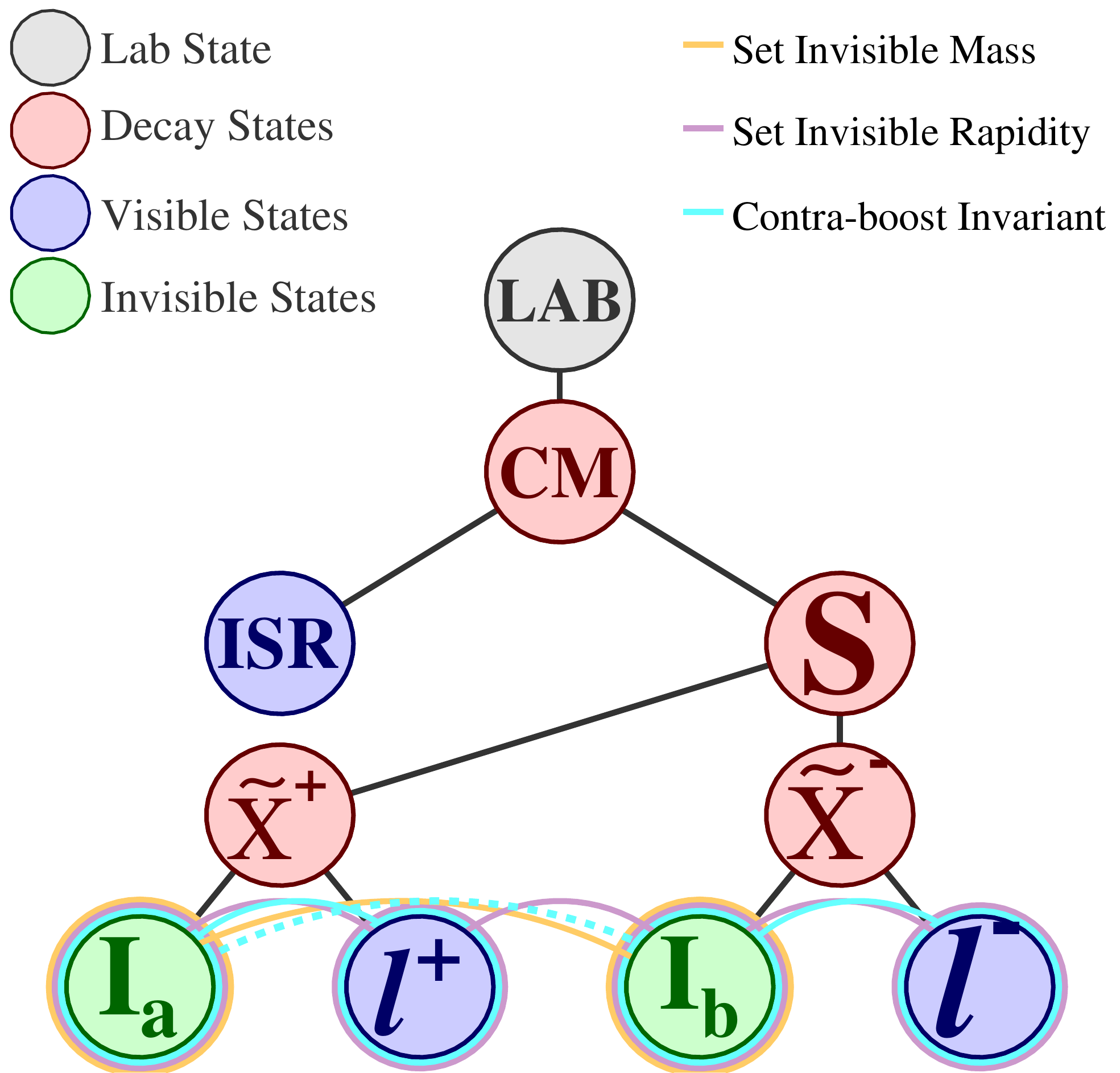}\caption{The decay tree for the analysis of compressed chargino pair production
in events with ISR. The substructure of the S system is specified
as follow: each chargino decays to a visible (lepton) and an invisible
(neutrino + neutralino) object. \label{fig:The-compressed-treeCC}}
\end{figure}

In order to improve the signal-to-background discrimination one enriches
the simplified version of the compressed RJR tree in Figure \ref{fig:The-compressed-decay},
by specifying the substructure of the S-system. This is feasible for 
the two leptons final state case when one has no ambiguity as to the provenance of the reconstructed
visible sparticle decay products. These decay products can then be assigned to the appropriate position in the tree.

The RJR decay tree is shown in Figure \ref{fig:The-compressed-treeCC}. Electrons
and muons are associated to the $l^{+}$ and $l^{-}$ systems, depending
to the electric charge, while the jets are assigned to the ISR-system. The S-system
frame is the approximation for the center-of-mass of the two charginos
and each one decays to a lepton and an invisible system. Each invisible
system collects the $\tilde{\chi}_{1}^{0}+\nu$ contribution of the hemisphere
$a$ and $b$. 

In this approach, a three dimensional view is considered and jigsaw
rules are applied in order to reconstruct the topology and the relevant
frames of reference. In the overall center-of-mass frame the ISR and S-systems
are back-to-back. A Lorentz invariant jigsaw rule is assumed for the estimate
of the mass of the invisible objects, while the rapidity is assigned
to the chargino center-of-mass (equal to the rapidity of the visible objects in the S-system). 
Finally, a contra-boost invariant jigsaw rule partitions the remaining unknown degrees
of freedom associated to I$_{a}$ and I$_{b}$. More information can
be found elsewhere \cite{JackProGen,JackProEW}. 

The useful transverse variables of the simplified tree can be computed
along with additional experimental observables. Having
in mind the simplified tree in Figure \ref{fig:The-compressed-decay},
one can reconstruct the I-system corresponding to the sum of the two
invisible systems $\mathrm{I}=\mathrm{I}_{a}+\mathrm{I}_{b}$ and
V to the sum of the two lepton systems $\mathrm{V}=l^{+}+l^{-}$ and compute
the transverse observables: $R_{\mathrm{ISR}}$, $p_{\mathrm{ISR},T}^{\mathrm{CM}}$
and $\Delta\phi_{\mathrm{ISR},\mathrm{I}}$.

Three-dimensional scale-sensitive variables and additional angular observables
include:
\begin{itemize}
\item $M^{\mathrm{V}}$ is the mass associated to the V-system: invariant
mass of ($l^{+}+l^{-}$).
\item $M^{\mathrm{\tilde{\chi}^{\pm}}}$ is the mass associated with the
chargino system.
\item $\Delta\phi_{\mathrm{l^{+}},\mathrm{I}}$ ($\Delta\phi_{\mathrm{l^{-}},\mathrm{I}}$)
polar angle between the positive (negative) charge lepton and ${\not\mathrel{\vec{E}}}_{T}$
computed in the Lab frame.
\item $\Delta\phi_{\mathrm{CM},\mathrm{I}}$: opening angle between the
CM system and the I-system.
\item $\cos\theta\equiv\hat{\beta}_{S}^{\mathrm{CM}}\cdot p_{\mathrm{I},T}^{\mathrm{S}}$:
the dot product between the direction of the boost from CM to the
reconstructed S-frame and the transverse momentum of the I-system
in the S-frame. 
\end{itemize}
Finally, jet multiplicities are considered. For the signal samples,
the mass associated to the chargino system $M^{\mathrm{\tilde{\chi}^{+}}}=M^{\mathrm{\tilde{\chi}^{-}}}$will
not reproduce the true chargino mass, since the true LSPs are massive and
the I$_{a,b}$ systems, in each hemisphere, are simplifications of the
neutralino plus neutrino contribution. Nevertheless, the distribution
of $M^{\mathrm{\tilde{\chi}^{\pm}}}$ is expected to be particularly
useful to distinguish the signal w.r.t. SM processes with similar
kinematics, but where lower mass objects populate one or both the $\tilde{\chi}^{\pm}$
systems. 

The main SM backgrounds are categorized into four groups:
1) Vector boson + jets, mainly populated by 
$Z\rightarrow \ell^{+}\ell^{-}$+jets; 2) Production of at least  
one top quark ($t$+X), with single-top and dileptonic $t\bar{t}$ both contributing; 
3) Irreducible di-boson processes mostly
arising from $W^{+}W^{-}$ with two leptons and missing transverse momentum; and 4) Contributions
such as vector boson fusion, tri-boson and gluon fusion plus jets with $h\rightarrow W^{+}W^{-}$,
are categorized as "others".
\begin{figure}[t]
\noindent \begin{centering}
\subfloat[\label{fig:MinvPresSF} ]{\noindent \begin{centering}
\includegraphics[height=5cm]{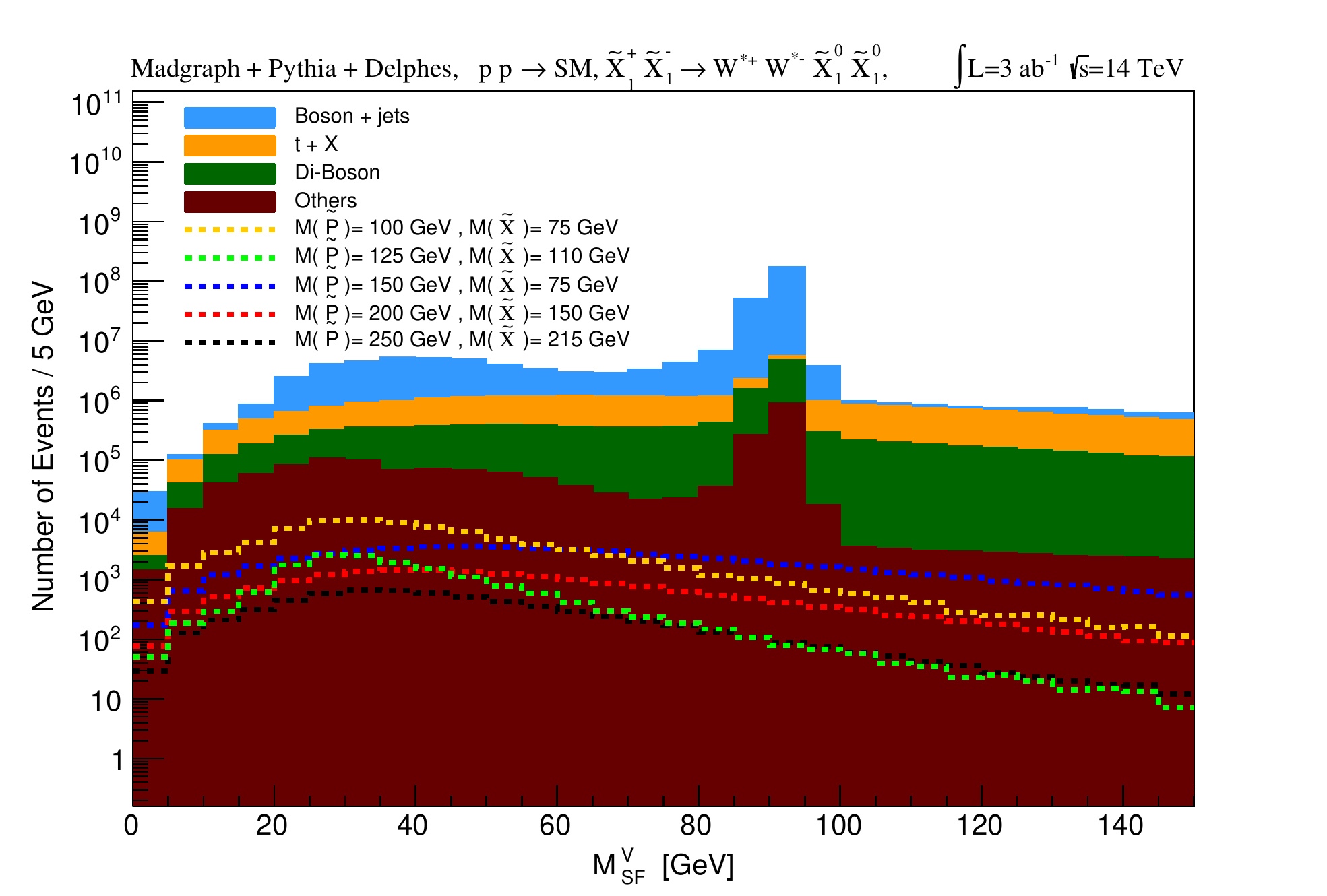}
\par\end{centering}

}$\;$\subfloat[\label{fig:MinvPresDF}]{\noindent \begin{centering}
\includegraphics[height=5cm]{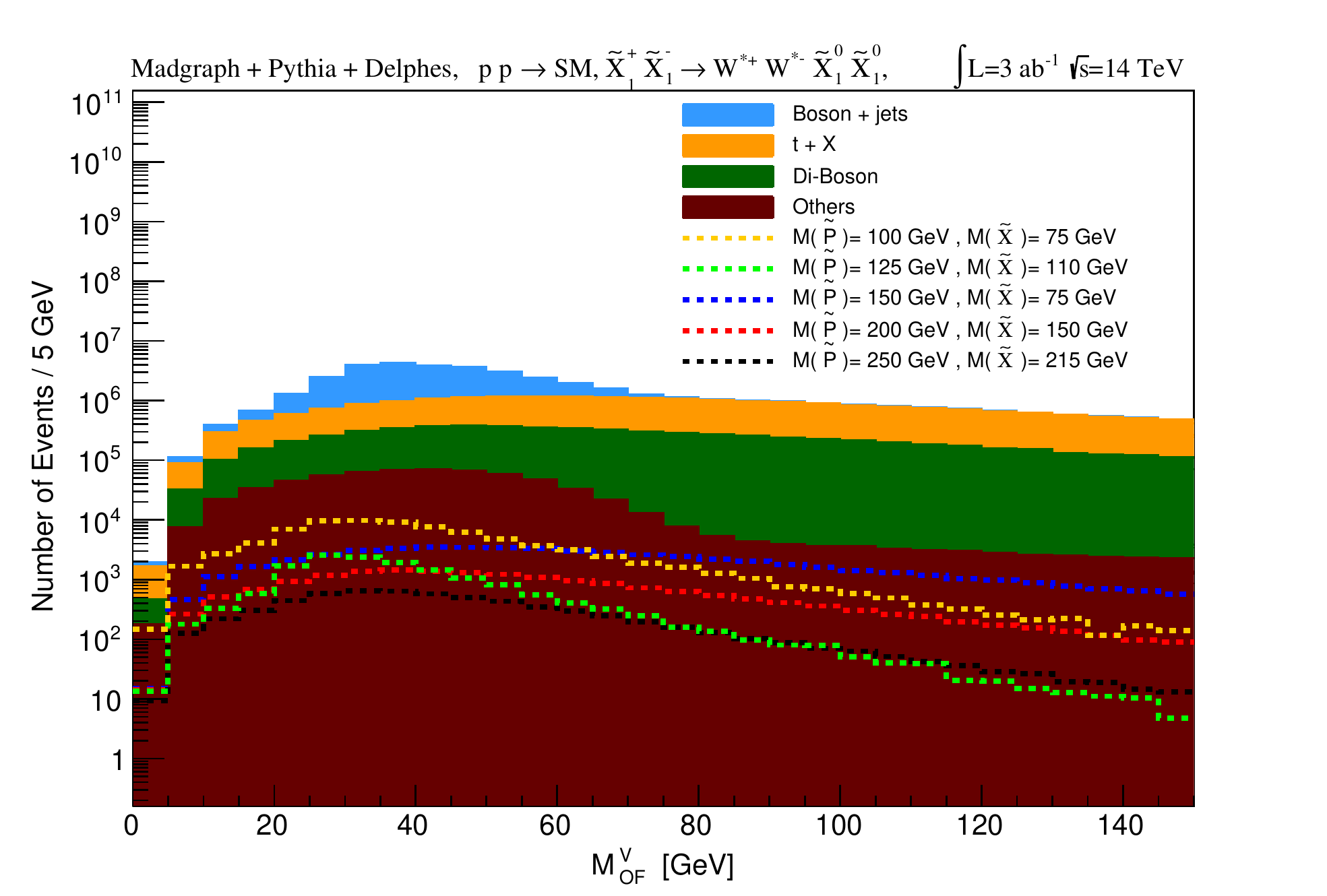}
\par\end{centering}

}
\par\end{centering}

\noindent \centering{}\caption{Distribution of the invariant mass of the two leptons for same (a)
or different (b) flavor.\label{fig:MinvPresSFDF} }
\end{figure}

Two leptons (electrons and muons) with $p_{T}>10$ GeV are required in the final state
and at least one jet with $p_{T}>20$ GeV, which
is associated to the ISR-system. Figure \ref{fig:MinvPresSFDF} shows
the distribution of the invariant mass of the two leptons for same
and different flavor, assuming a minimal value for the transverse
missing momentum ${\not\mathrel{E}}_{T}>20$ GeV. Standard Model background
samples are stacked together, while the overlaid dashed curves refer
to chargino pair production samples with different masses and mass
splittings. Notice the peak around 90 GeV for same flavor leptons,
due to the Standard Model backgrounds containing $Z$ bosons produced in association with jets (in blue), with a moderate contribution from
$WZ$ (in green) and vector boson fusion and tri-boson (in red).
In the compressed regime the final state events for all the signal
distributions tend to populate lower values of $M^{\mathrm{V}}$ and
a requirement $M^{\mathrm{V}}<70$ GeV, or tighter, will be
used to specify the signal regions. Notice the additional peak for
low values of $M^{\mathrm{V}}$, arising from $Z$+ jets and vector boson
fusion contributions resulting in a comparable number of events for
the cases with two leptons with same or different flavor. The main
processes that contribute in this region are $Z\rightarrow\tau^{+}\tau^{-}\rightarrow l^{+}l^{-}\nu\nu\nu\nu$,
and moderate contribution from Drell-Yan processes with missing transverse momentum
($Z^{*}(\gamma*)\rightarrow\tau^{+}\tau^{-}$) or $W$
boson production, decaying leptonically, with an additional lepton faked by
a jet or a photon. For the di-leptonic decay of the $Z$ boson via
taus the value for $M^{V}$ is reconstructed to be below the $Z$ mass, 
representing a challenge to the analysis in search of compressed charginos.

Herein, we consider the preselection criteria as follows: final states
with two leptons and at least one light jet. A veto is applied for
the jets tagged as $b$, $\tau$ and fat: $N_{b-\mathrm{jet}}^{\mathrm{ISR}}=0$ and $N_{\tau-\mathrm{jet}}^{\mathrm{ISR}}=0$
and $N_{\mathrm{fat}}^{\mathrm{ISR}}=0$.\footnote{In this work a fat jet is defined with $M>60$ GeV and is a candidate
for boosted SM Higgs, vector bosons and top-quark decaying hadronically.} A minimal value for the
missing transverse momentum (${\not\mathrel{E}}_{T}>50$ GeV) in concert
with $p_{\mathrm{ISR},T}^{\mathrm{CM}}$> 50 GeV is required. In addition,
the criterion $M^{\mathrm{V}}<70$ GeV is imposed. This requirement
excludes a large portion of the Standard Model background events,
in particular $t\bar{t}$ and multi-bosons processes, independently
from the flavor of the two leptons reconstructed. Standard Model
processes involving a meson decaying in two same flavor leptons are
expected with a small value of the invariant mass $\lesssim10$ GeV:
notably, signal sample events tend to assume larger values. 

We present the impact of the main RJR observables sensitive
to probe compressed chargino pair mass spectra to reduce the specific
Standard Model contribution and progressively, the related selection
criteria will be imposed.

Numerous Standard Model processes result in a low
value of $M^{\mathrm{V}}$, in particular the boson plus jets contribution.
The focus is on the process $Z\rightarrow\tau^{+}\tau^{-}\rightarrow l^{+}l^{-}\nu\nu\nu\nu$
plus jets. For such events the role of the chargino system in Figure
\ref{fig:The-compressed-treeCC} is assumed by the tau's leptonic decay,
while the $\mathrm{I}$ systems reconstruct the information of the
two neutrinos in each hemisphere. For these background events $M^{\mathrm{\tilde{\chi}^{\pm}}}$
is a reconstruction for the mass of the lepton and two neutrinos resulting
from the $\tau$ decays.
\begin{figure}[t]
\noindent \begin{centering}
\subfloat[\label{fig:2DCCRMSDiBoson-1-1} ]{\noindent \begin{centering}
\includegraphics[width=0.32\textwidth]{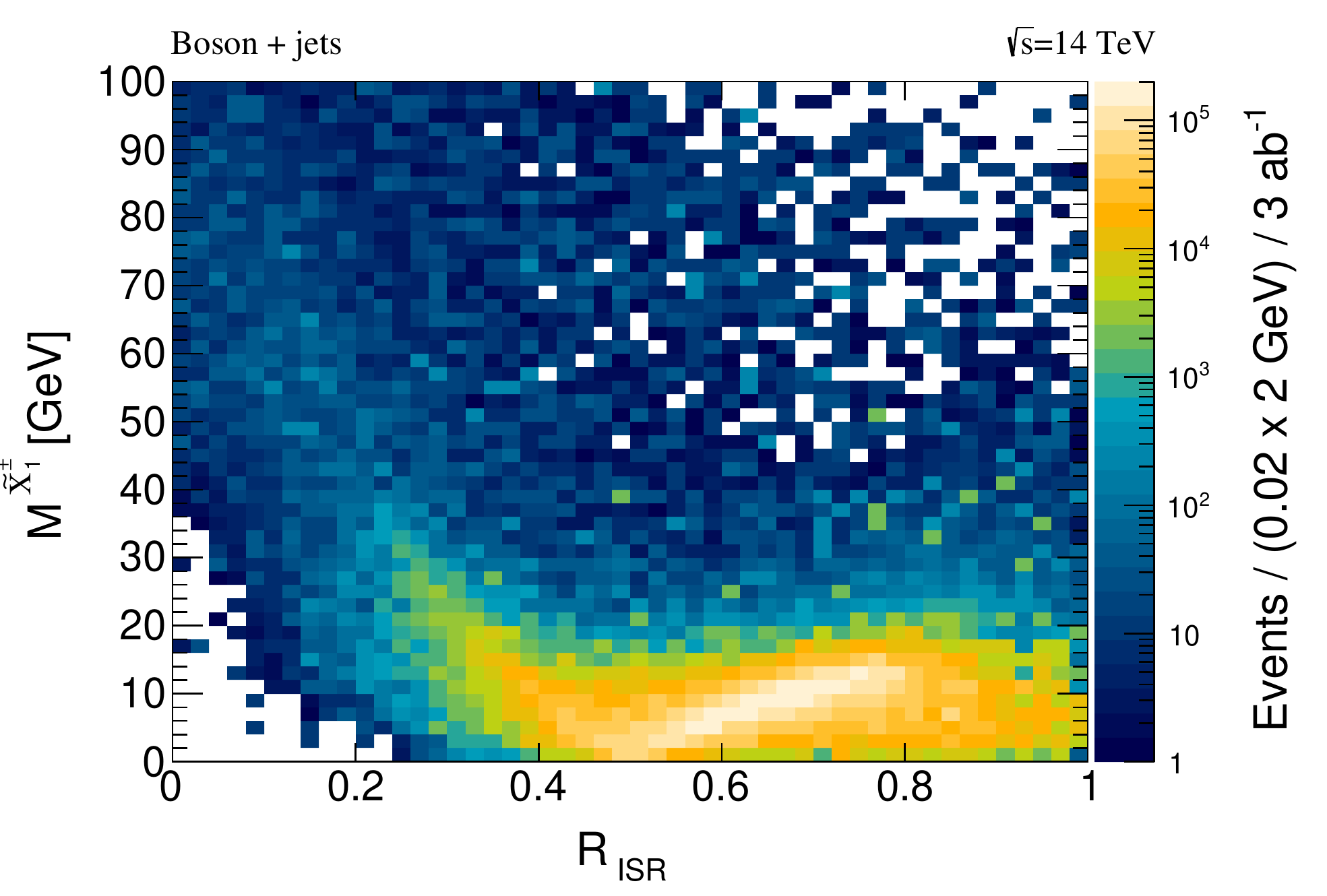}
\par\end{centering}

}\subfloat[\label{fig:2DCCRMS200-1-1} ]{\noindent \begin{centering}
\includegraphics[width=0.32\textwidth]{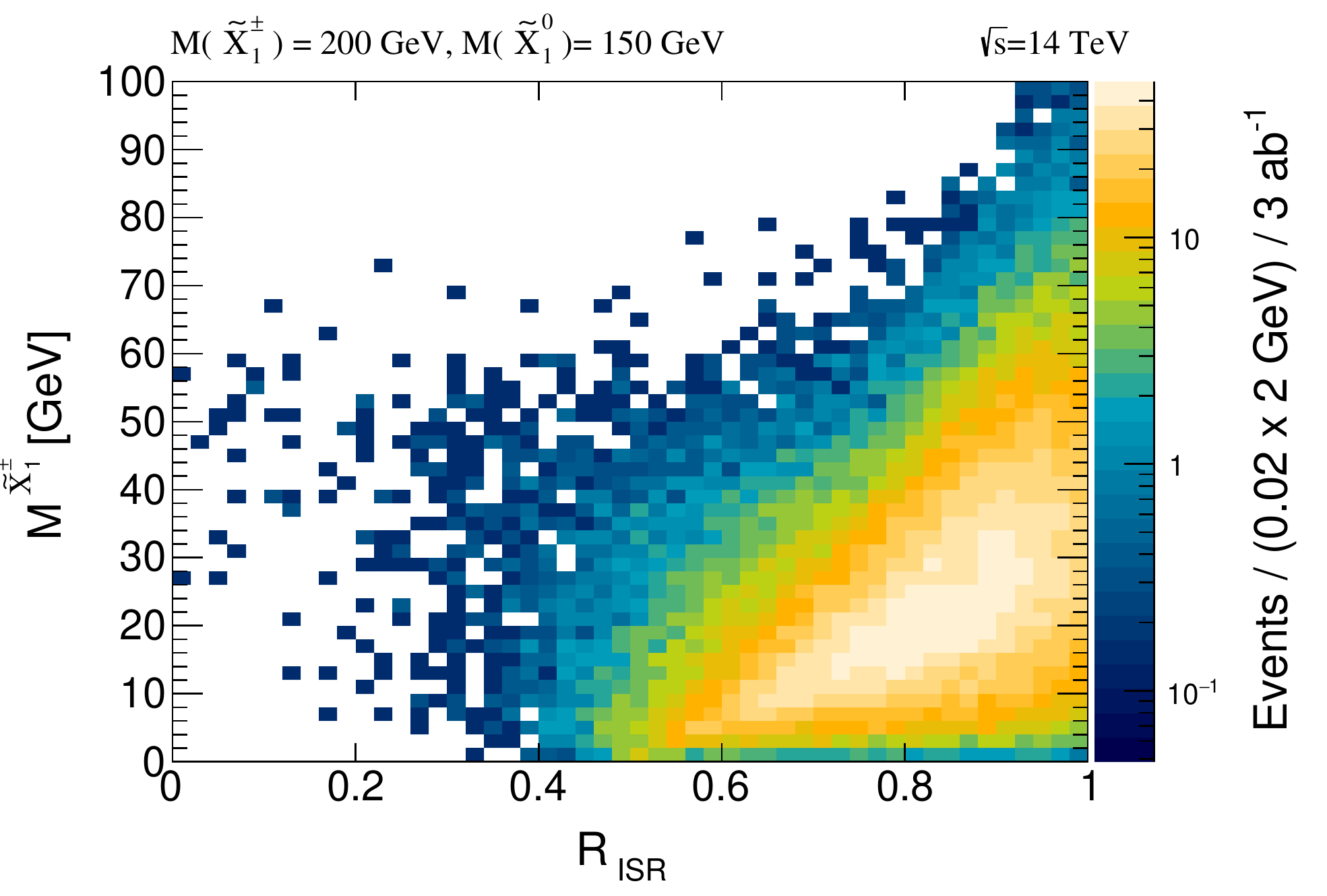}
\par\end{centering}

}\subfloat[\label{fig:1DMCharg} ]{\noindent \begin{centering}
\includegraphics[width=0.33\textwidth]{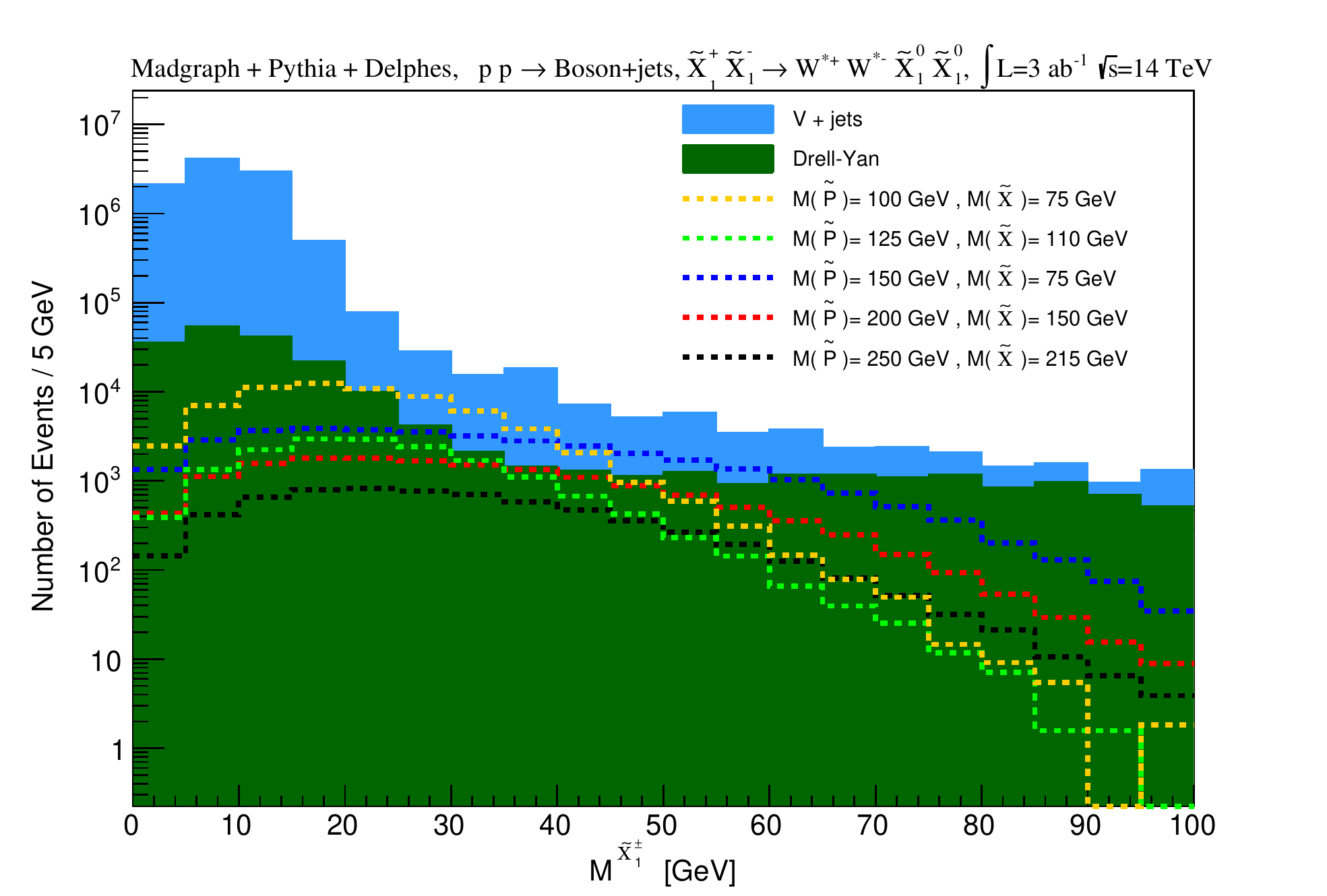}
\par\end{centering}

}
\par\end{centering}

\noindent \centering{}\caption{Two-dimensional distribution of $M^{\mathrm{\tilde{\chi}^{\pm}}}$
as a function of $R_{\mathrm{ISR}}$ for the Standard Model $V+$jets
background (a) and the signal samples
$M_{\tilde{\chi}_{1}^{\pm}}$=200 GeV, $M_{\tilde{\chi}_{1}^{0}}$=150
GeV (b) and distribution of $M^{\mathrm{\tilde{\chi}^{\pm}}}$
(c) for the events expected per bin for an integrated
luminosity of 3 $\mathrm{ab}^{-1}$ at $\sqrt{s}=14$ TeV satisfying
the preselection criteria. \label{fig:2DMSRCC-1-1}}
\end{figure}
 The first two plots in Figure \ref{fig:2DMSRCC-1-1} show the two-dimensional
distributions between $M^{\mathrm{\tilde{\chi}^{\pm}}}$and the ratio
$R_{\mathrm{ISR}}$ for the boson plus jets backgrounds and the signal
sample $M_{\tilde{\chi}_{1}^{\pm}}$= 200 GeV and $M_{\tilde{\chi}_{1}^{0}}$=
150 GeV, Figure \ref{fig:1DMCharg} shows the distribution of $M^{\mathrm{\tilde{\chi}^{\pm}}}$
for the five signal samples and for the on-shell/off-shell, boson
plus jets backgrounds. We demand $M^{\mathrm{\tilde{\chi}^{\pm}}}$>
24 GeV in order to suppress the V+jets background. 

\begin{figure}[t]
\noindent \begin{centering}
\subfloat[\label{fig:2DCCRNjtt} ]{\noindent \begin{centering}
\includegraphics[width=0.32\textwidth]{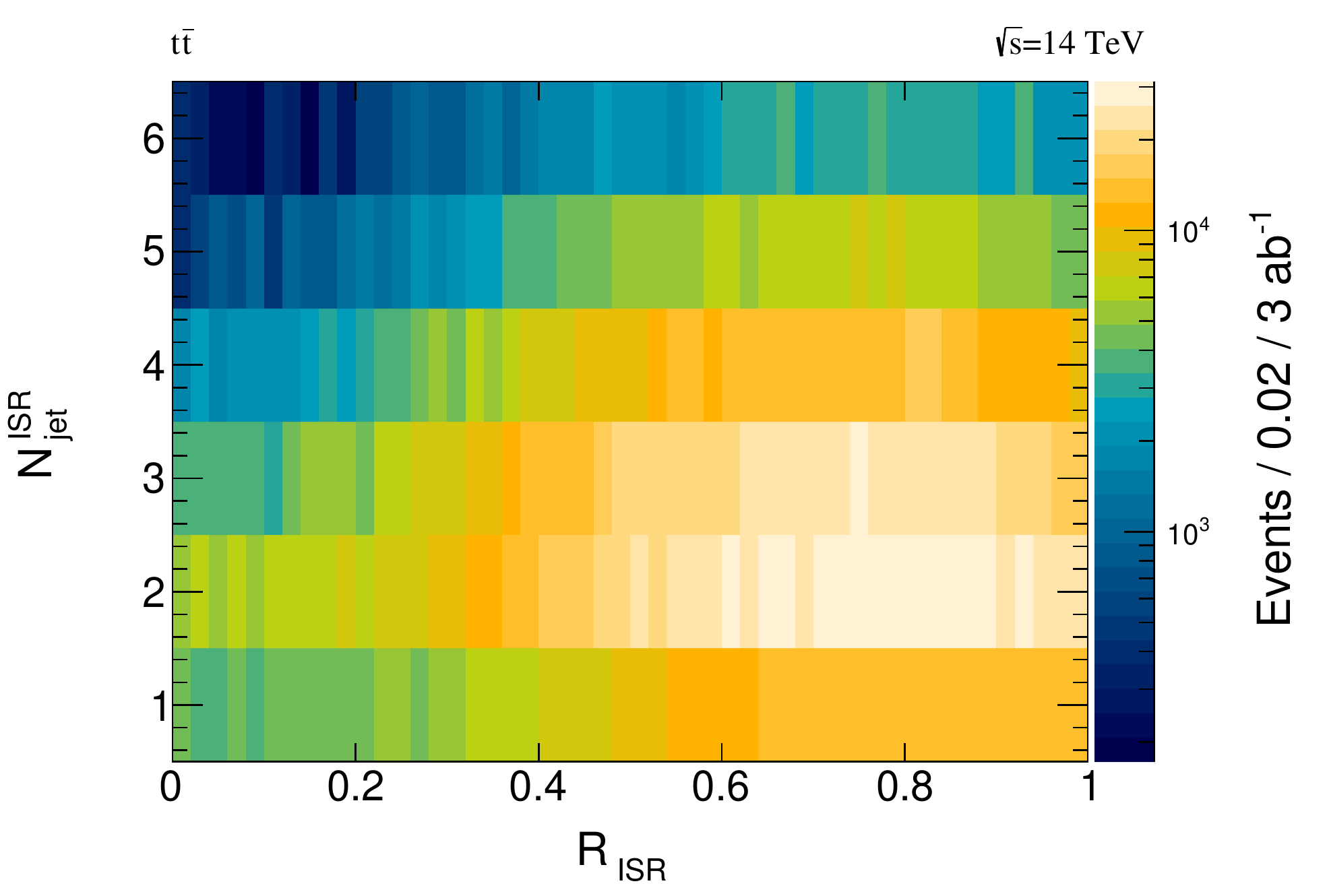}
\par\end{centering}

}\subfloat[\label{fig:2DCCRNjtt-1} ]{\noindent \begin{centering}
\includegraphics[width=0.32\textwidth]{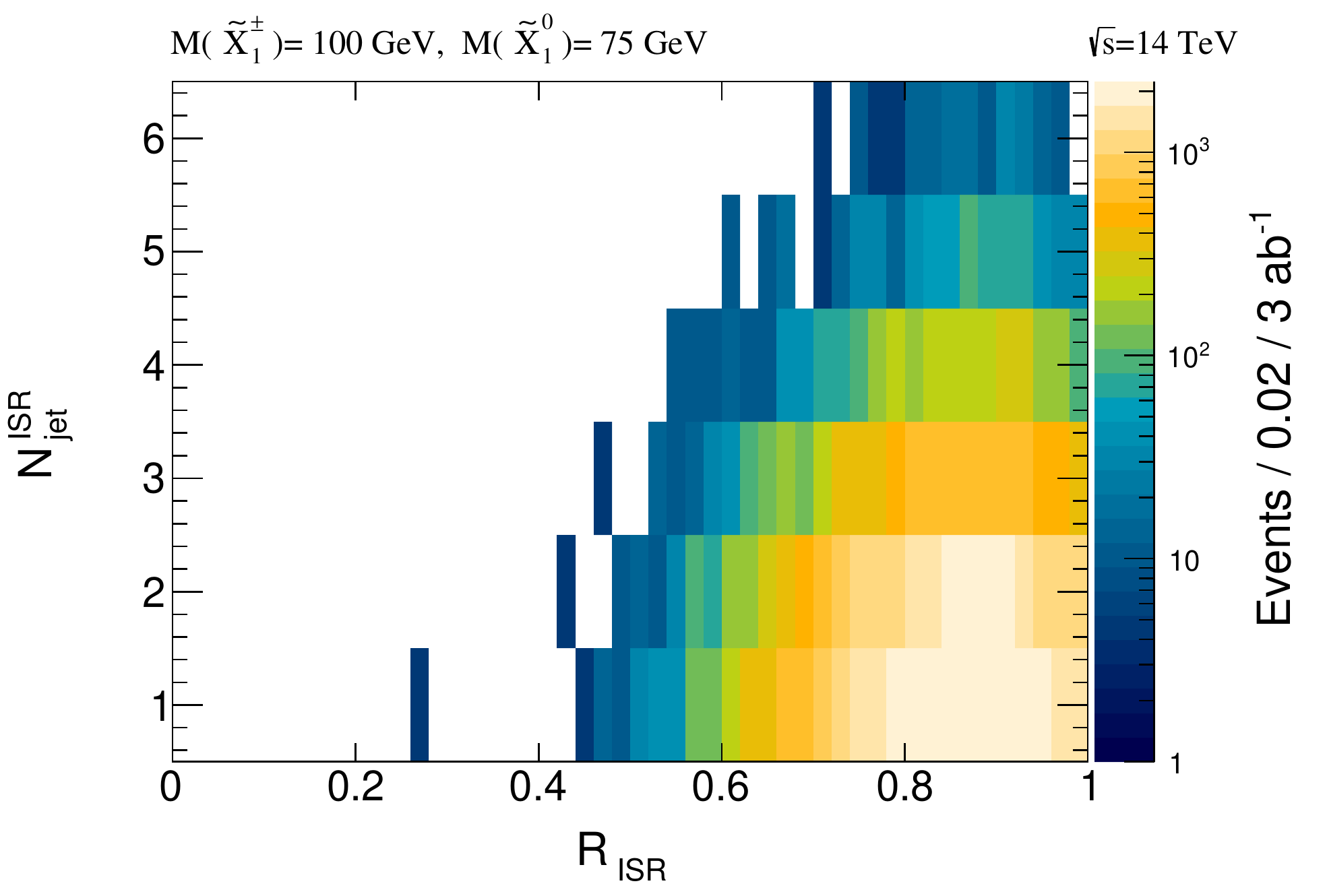}
\par\end{centering}

}\subfloat[\label{fig:2DCCRNjtt-2} ]{\noindent \begin{centering}
\includegraphics[width=0.32\textwidth]{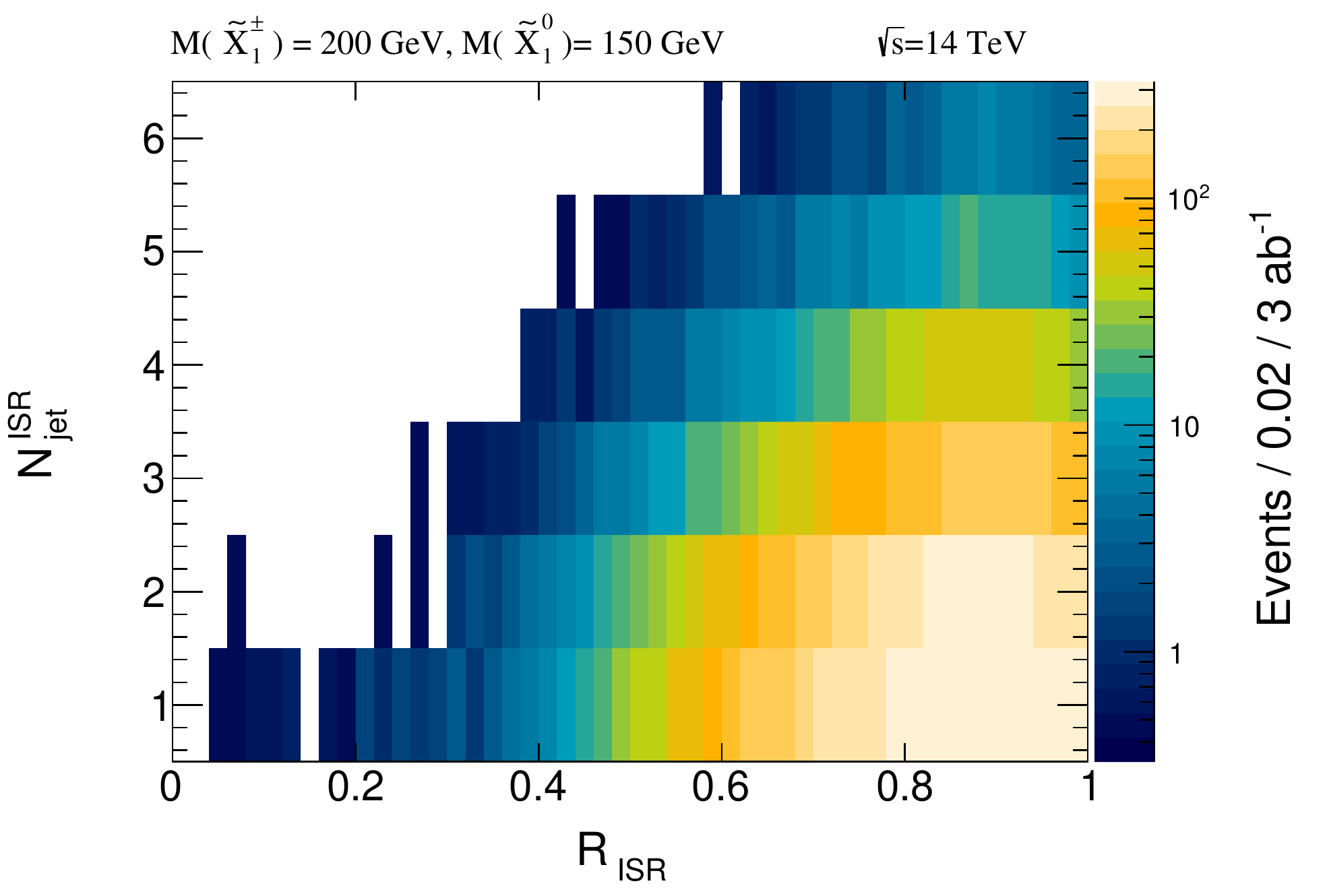}
\par\end{centering}

}
\par\end{centering}

\noindent \centering{}\caption{Distribution of $N_{\mathrm{jet}}^{\mathrm{ISR}}$ as a function of
$R_{\mathrm{ISR}}$ for the Standard Model $t\bar{t}$ background
(a) and the signal samples $M_{\tilde{\chi}_{1}^{\pm}}$=100,
200 GeV and $M_{\tilde{\chi}_{1}^{0}}$=75, 150 GeV in (b and c). We impose preselection criteria and
$M^{\mathrm{\tilde{\chi}^{\pm}}}$> 24 GeV.\label{fig:TwoDCCNjRttbarsig}}
\end{figure}

With this requirement the SM background is dominated by top processes,
specifically a pair of (on- or off-shell) top quarks in the di-leptonic
channel. Figure \ref{fig:TwoDCCNjRttbarsig} shows the distribution
of the light jet multiplicity as a function of the ratio. 
In order to attenuate the $t\bar{t}$ contribution we demand only
one jet in the final state. Despite the $N_{\mathrm{jet}}^{\mathrm{ISR}}=1$
requirement, and vetoing on jets coming from the fragmentation of
bottoms, the $t\bar{t}$ background is still not suppressed. If the
requirement $N_{\mathrm{fat}}^{\mathrm{ISR}}=0$ attenuates the contribution
with the two jets reconstructed in similar directions, one of the two jets
could be outside the geometrical acceptance, mis-measured or of too low momentum to be reconstructed.
Also if these events are relatively rare, their contribution is not
negligible due to their high cross section $\sigma_{pp\rightarrow t\bar{t}}\sim$$\mathcal{O}$($10^{3}$
pb) at 14~TeV LHC collisions.

\begin{figure}[t]
\noindent \begin{centering}
\subfloat[\label{fig:CCdphiL+I} ]{\noindent \begin{centering}
\includegraphics[width=0.33\textwidth]{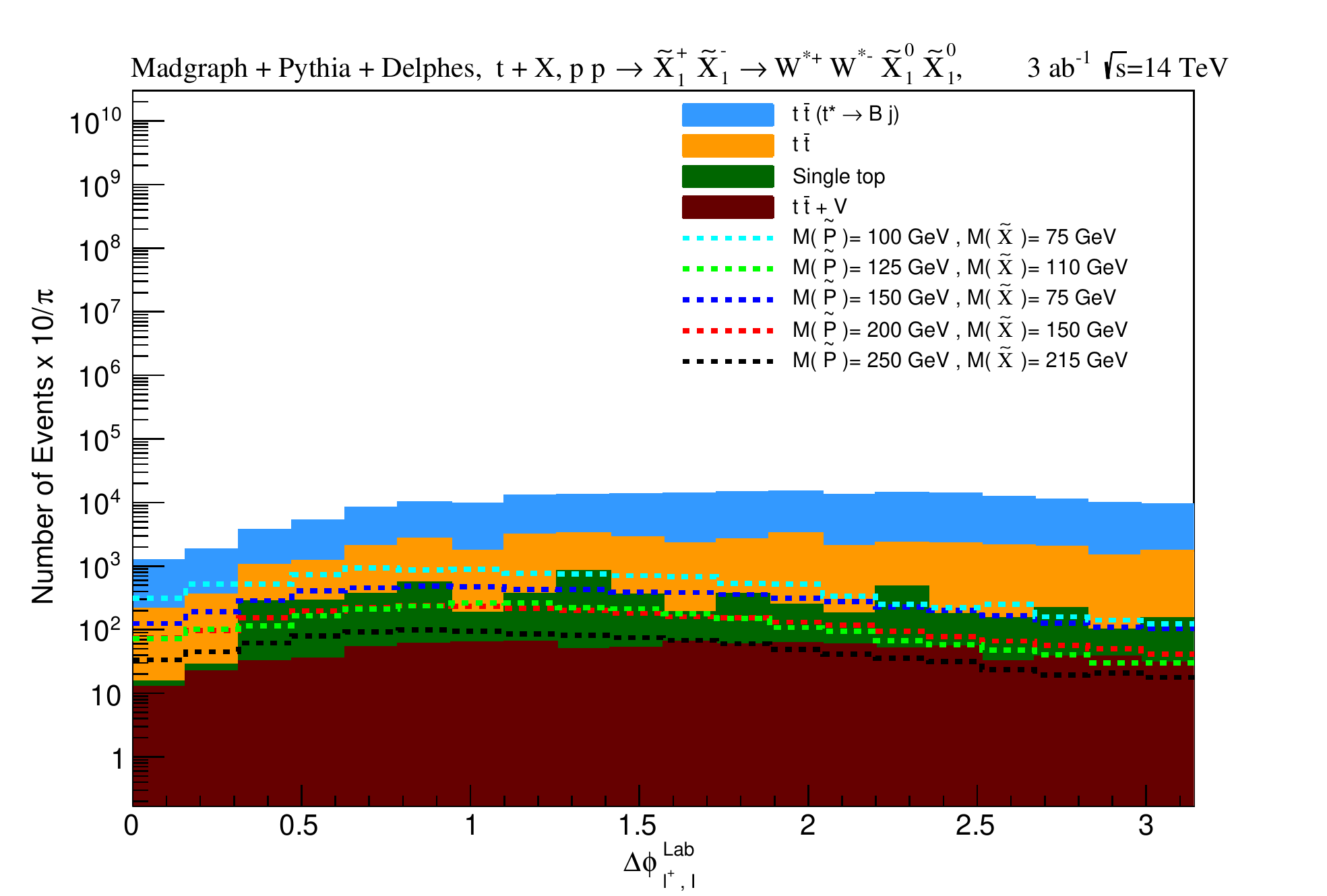}
\par\end{centering}

}\subfloat[\label{fig:CCdphiL1dphiL2} ]{\noindent \begin{centering}
\includegraphics[width=0.32\textwidth]{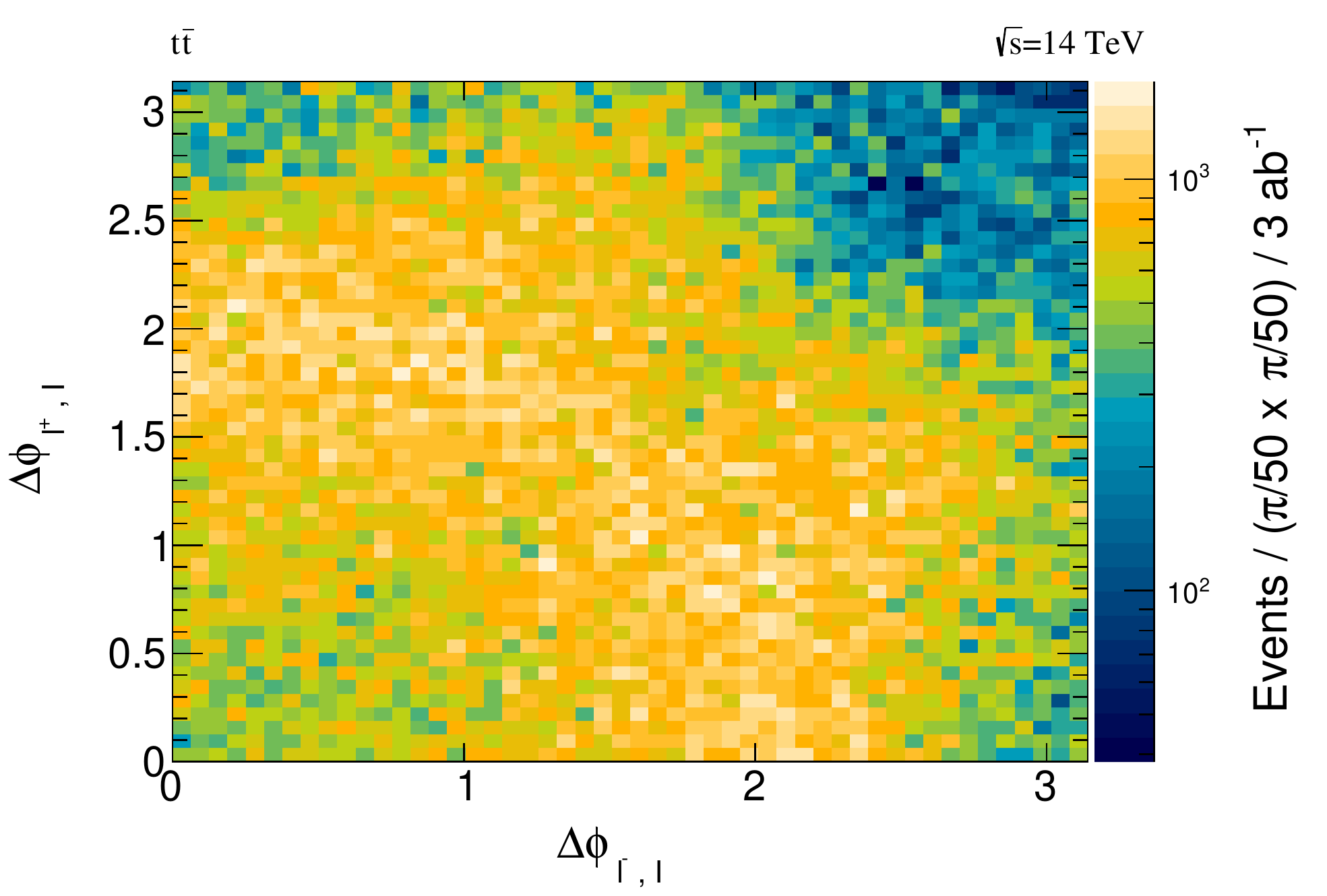}
\par\end{centering}

}\subfloat[\label{fig:CCdphiL+IdphiL-} ]{\noindent \begin{centering}
\includegraphics[width=0.32\textwidth]{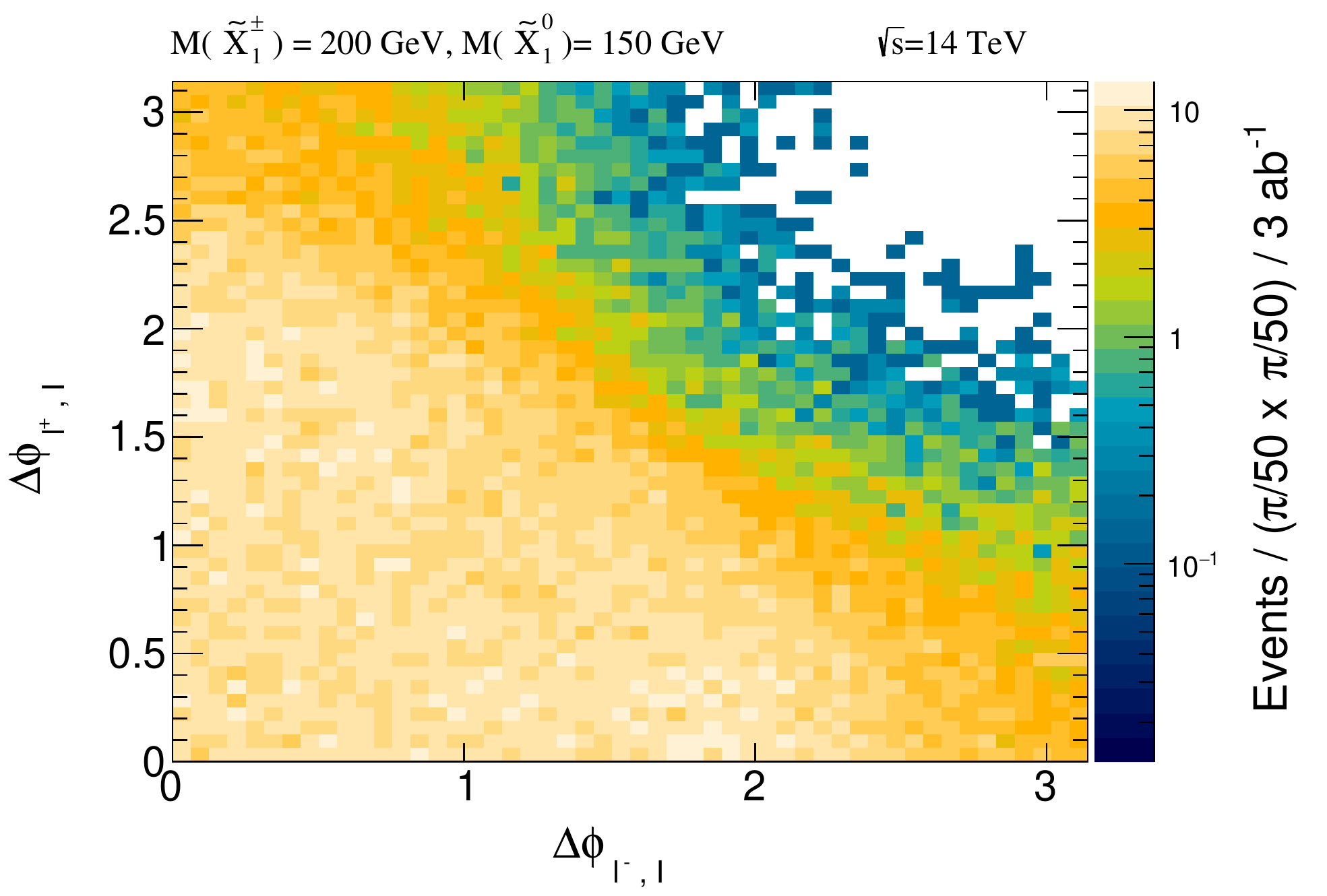}
\par\end{centering}

}
\par\end{centering}

\noindent \centering{}\caption{Distribution of $\Delta\phi_{\mathrm{l^{+}},\mathrm{I}}$ for Standard
Model $t$ + X background and signal events passing preselection criteria
and the additional requirements $M^{\mathrm{\tilde{\chi}^{\pm}}}$>
24 GeV and $N_{\mathrm{jet}}^{\mathrm{ISR}}=1$ (a).
Two-dimensional distribution of the opening angles between the leptons
and the I-system for the $t\bar{t}$ background (b)
and the signal sample $M_{\tilde{\chi}_{1}^{\pm}}$= 200 GeV, $M_{\tilde{\chi}_{1}^{0}}$=
150 GeV (c) imposing $R_{\mathrm{ISR}}>0.6$.
\label{fig:dphil+l- all}}
\end{figure}

Figure \ref{fig:CCdphiL+I} shows the distribution of $\Delta\phi_{\mathrm{l^{+}},\mathrm{I}}$
for the signal samples and the $t$ + X backgrounds categorized in
four sub-processes and stacked together. The events from the top pair
contributions tend to populate value close to $\pi$, while signal-like
events populate low values. Figures \ref{fig:CCdphiL1dphiL2} and \ref{fig:CCdphiL+IdphiL-}
show the two-dimensional distribution $\Delta\phi_{\mathrm{l^{+}},\mathrm{I}}$
vs $\Delta\phi_{\mathrm{l^{-}},\mathrm{I}}$ for the $t\bar{t}$ background
and the signal sample $M_{\tilde{\chi}_{1}^{\pm}}$=200 GeV and $M_{\tilde{\chi}_{1}^{0}}$=150
GeV, assuming the same selection criteria and requiring $R_{\mathrm{ISR}}>0.6$.
The requirements select background events with kinematics
similar to the signal events and in particular we see that a simultaneously large value 
of $\Delta\phi_{\mathrm{l^{\pm}},\mathrm{I}}$ for both
the leptons is disfavored. Such events contain predominantly two top quarks
produced with low transverse momenta resulting in final states with
two reconstructed leptons and one jet not properly tagged. In the
transverse plane, one of the two leptons is expected to fly close
to the reconstructed jet (associated to the ISR-system), while the
other is expected to be closer to the invisible system. As a consequence
background events tend to assume larger values of $\Delta\phi_{\mathrm{l^{+}},\mathrm{I}}+\Delta\phi_{\mathrm{l^{-}},\mathrm{I}}$
than signal events. A similar two-dimensional distribution as in
Figure \ref{fig:CCdphiL+IdphiL-} is demonstrated by all signal samples studied.
To attenuate the $t+$X background one requires a unique light jet
associated to the ISR-system and in addition $\Delta\phi_{\mathrm{l^{+}},\mathrm{I}}+\Delta\phi_{\mathrm{l^{-}},\mathrm{I}}<2$.

Applying these selection criteria, the dominant Standard Model 
contribution is the irreducible di-boson background: $W^{+}W^{-}.$ The goal
is to distinguish between signal and background events with similar
event topologies and kinematics, in particular when selection criteria close
to the final configuration are applied. The key difference to exploit is that of the I-system ($\mathrm{I}_{a}+\mathrm{I}_{b}$) for
the $W^{+}W^{-}$ background composed of two neutrinos, while
the signal events have four weakly interacting particles comprising the invisible system. 

\begin{figure}[t]
\noindent \begin{centering}
\subfloat[\label{fig:dphiISRIWW} ]{\noindent \begin{centering}
\includegraphics[width=0.32\textwidth]{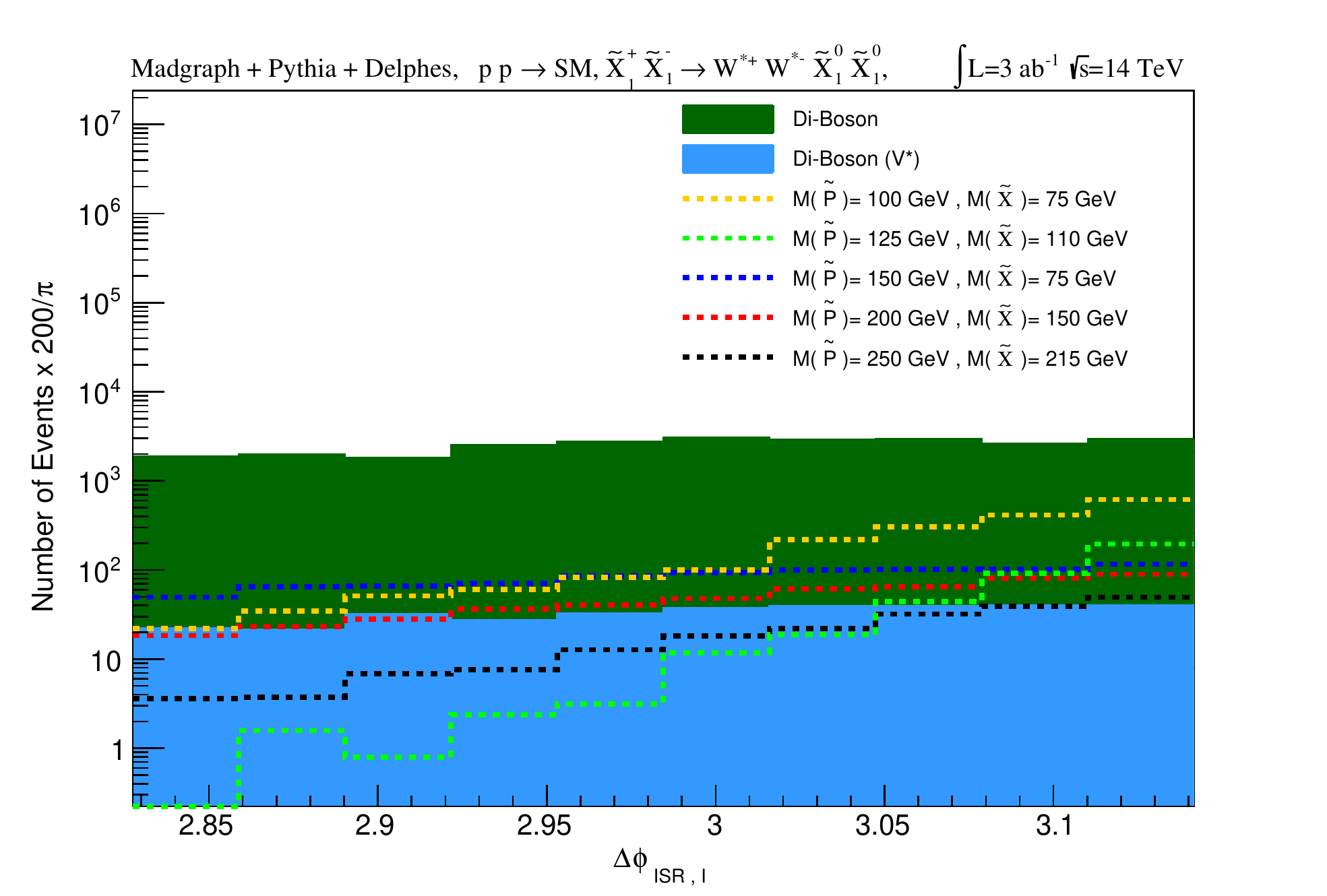}
\par\end{centering}

}\subfloat[\label{fig:dphiCMIWW} ]{\noindent \begin{centering}
\includegraphics[width=0.32\textwidth]{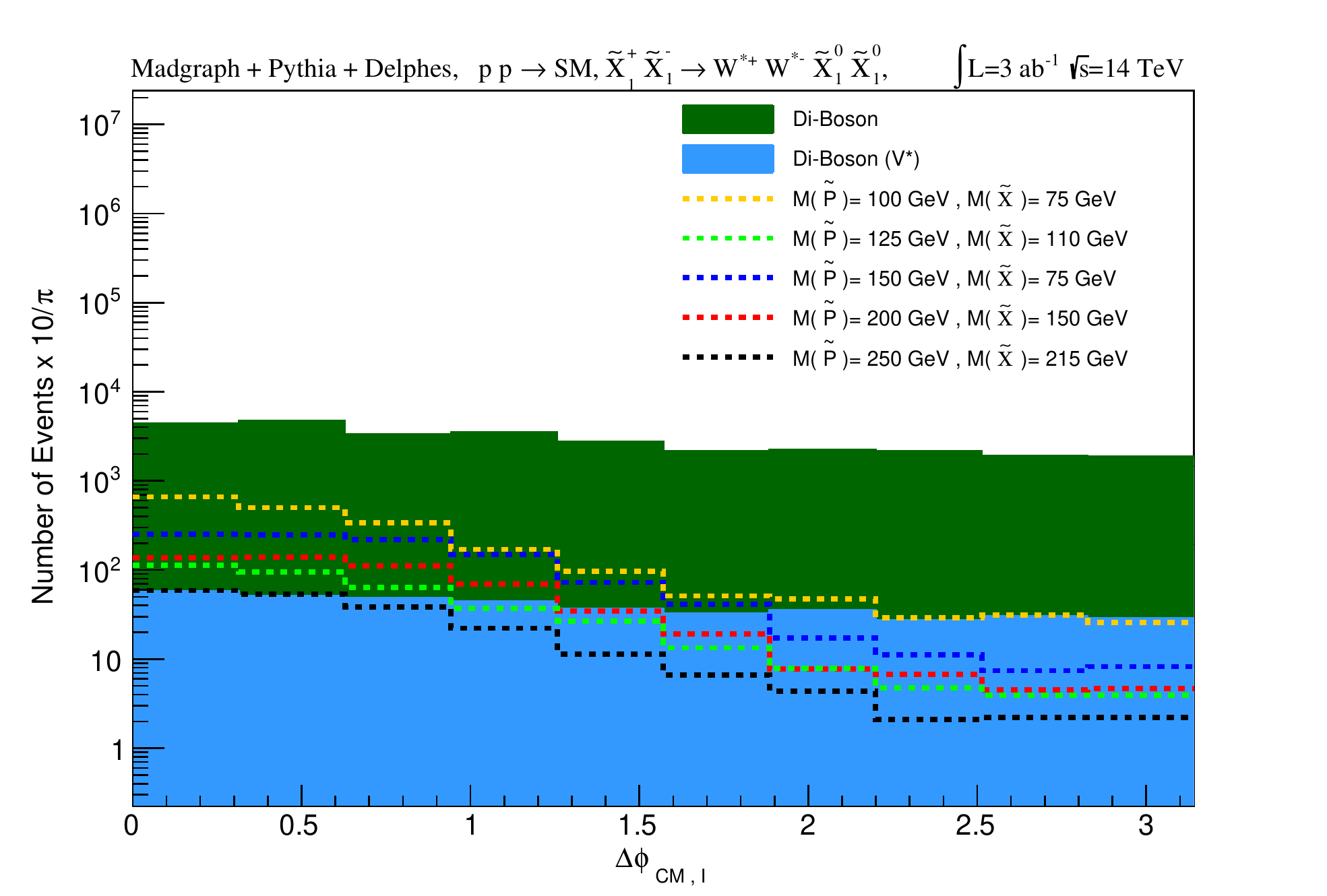}
\par\end{centering}

}\subfloat[\label{fig:CCcosSWW} ]{\noindent \begin{centering}
\includegraphics[width=0.32\textwidth]{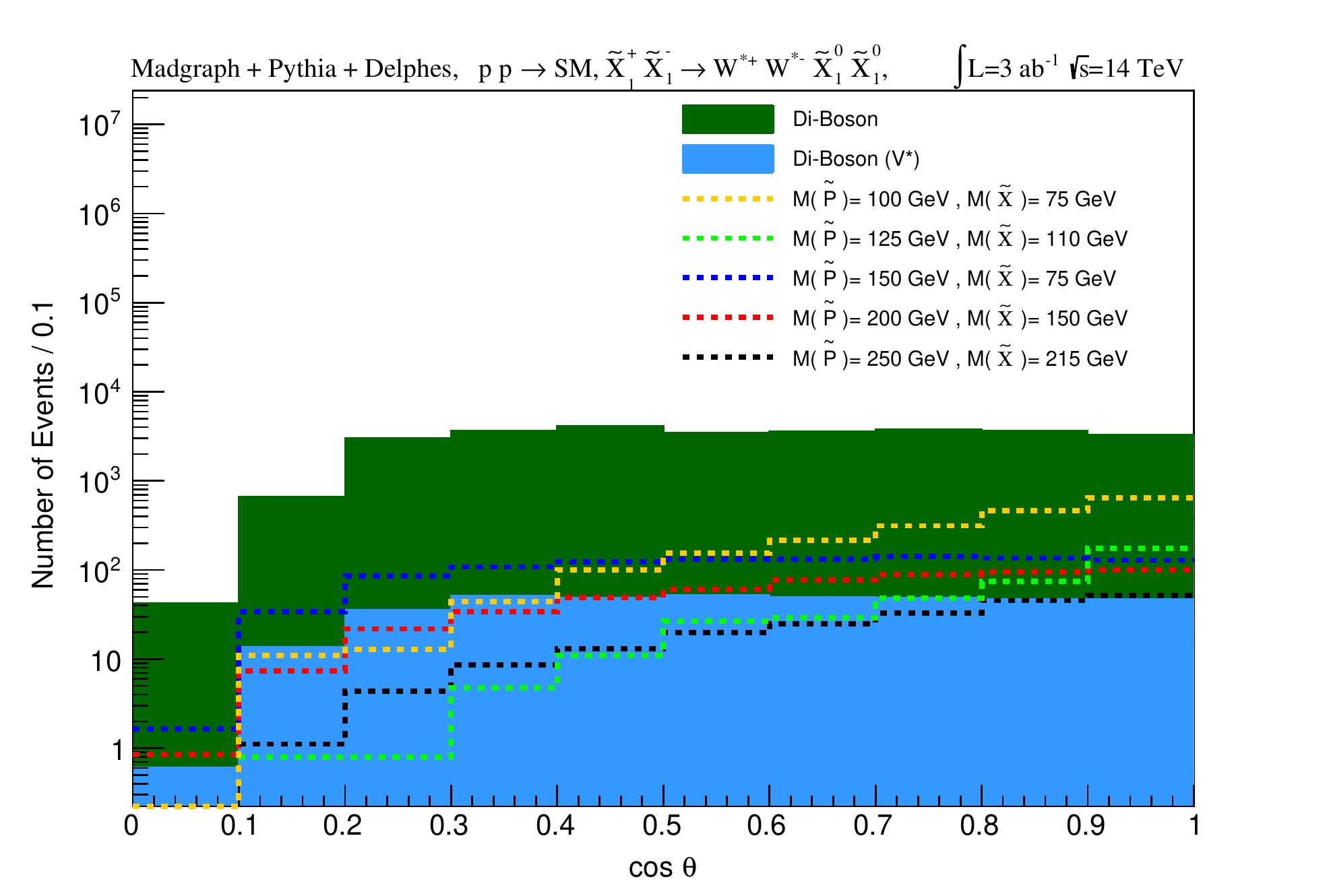}
\par\end{centering}

}
\par\end{centering}

\noindent \centering{}\caption{Distributions of angular RJR observables for signal samples and di-boson
Standard Model background, for events passing preselection criteria
and the additional requirements $M^{\mathrm{\tilde{\chi}^{\pm}}}$>
24 GeV, $N_{\mathrm{jet}}^{\mathrm{ISR}}=1$, $R_{\mathrm{ISR}}>0.6$
and $\Delta\phi_{\mathrm{l^{+}},\mathrm{I}}+\Delta\phi_{\mathrm{l^{-}},\mathrm{I}}<2$.
\label{fig:CC1DRdphidphicoss}}
\end{figure}

Figure \ref{fig:CC1DRdphidphicoss} shows the main angular observables
sensitive to separate events resulting from compressed chargino samples
with respect to $WW$ decays. Figure \ref{fig:dphiISRIWW}
shows the distribution of $\Delta\phi_{\mathrm{ISR},\mathrm{I}}$.
Signal events tend to populate values closer to $\pi$, as the 
the mass difference $\Delta M=M_{\tilde{P}}-M_{\tilde{\chi}}$ is reduced. Figure
\ref{fig:dphiCMIWW} shows the distribution of the angle between the
CM and I-system, where in this case signal events towards zero, almost independently of $\Delta M$
or $M_{\tilde{P}}/M_{\tilde{\chi}}$. The distribution of
$\cos\theta\equiv\hat{\beta}_{S}^{\mathrm{CM}}\cdot p_{\mathrm{I},T}^{\mathrm{S}}$
is shown in Figure \ref{fig:CCcosSWW}. 
\begin{comment}
%...START maybe cut or simpler...

For nearly degenerate parent-child superparticles, the direction of
the LSP is roughly the same of the sparent one. For the signal, the
sum of the transverse momenta of the two charginos is the transverse
momentum of the S-system, while the main contribution to the direction
of the transverse momentum of the I-system is given by the two LSPs:
each neutrino has an angular separation respect the associated LSP
direction, but the sum of the two contributions is expected to be
zero in average. In the transverse plane the resulting direction of
the V-system is so expected to be very close to the I-system resulting
from four particles. For the background instead the angular separation
of each neutrino respect the original W direction is distributed much
more democratically. The resulting I direction can differ considerably
respect the S direction of the two \textit{W}s, providing in such
a case a larger separation with the V-system in the transverse plane.
The same behavior must occur in three dimensions. The observable $\cos\theta$
is expected to be sensitive to the angular separation between the
V=$l^{+}+l^{-}$ and the I=$I_{a}+I_{b}$ system assuming the z-momenum
of the I-system to be zero in the S frame.

%...END maybe cut or simpler...
\end{comment}

Selection criteria defined with the compressed RJR observables result
in signal regions used to investigate chargino pair production in
final states with two leptons and missing transverse momentum. The
requirements for the observable $R_{\mathrm{ISR}}$ are tuned depending
on the mass ratio and are more stringent than the chargino-neutralino
associated study due to the larger multiplicity of weakly interacting
particles in the final state.
\begin{table}[tp]
\centering{}{\small{}}%
\begin{tabular}{|>{\centering}p{0.26\textwidth}|>{\centering}p{0.11\textwidth}|>{\centering}p{0.11\textwidth}|>{\centering}p{0.11\textwidth}|>{\centering}p{0.11\textwidth}|>{\centering}p{0.11\textwidth}|}
\cline{2-6} 
\multicolumn{1}{>{\centering}p{0.26\textwidth}|}{\foreignlanguage{british}{}} & \multicolumn{5}{c|}{{\small{}$\begin{array}{c}
\\
\\
\end{array}$}\textbf{Mass Splitting {[}GeV{]}}{\small{}$\begin{array}{c}
\\
\\
\end{array}$}}\tabularnewline
\hline 
{\small{}$\begin{array}{c}
\\
\\
\end{array}$}\textbf{Variable}{\small{} $\begin{array}{c}
\\
\\
\end{array}$} & {\small{}$\Delta M=15$} & {\small{}$\Delta M=25$} & {\small{}$\Delta M=35$} & {\small{}$\Delta M=50$} & {\small{}$\Delta M=75$}\tabularnewline
\hline 
\multicolumn{6}{c}{
\vspace{-0.4cm}
\selectlanguage{english}%
}\tabularnewline
\hline 
{\small{}Object multiplicity} & \multicolumn{5}{c|}{{\small{}2 OS Leptons ($e$ and $\mu$) with $p_{T}^{lep}>10$ GeV, }}\tabularnewline
{\small{}selection criteria} & \multicolumn{5}{c|}{{\small{}$N_{jet}^{\mathrm{ISR}}=1$, $N_{b-\mathrm{jet}}^{\mathrm{ISR}}=0,$
$N_{\tau-\mathrm{jet}}^{\mathrm{ISR}}=0$ and $N_{\mathrm{fat}}^{\mathrm{ISR}}=0$ }}\tabularnewline
\hline 
{\small{}$p_{\mathrm{ISR},T}^{\mathrm{CM}}$ (${\not\mathrel{E}}_{T}$)$>$
{[}GeV{]}} & \multicolumn{5}{c|}{{\small{}$50$}}\tabularnewline
\hline 
{\small{}$M^{\mathrm{V}}<$ {[}GeV{]}} & \multicolumn{3}{c|}{{\small{}$50$}} & {\small{}$60$} & {\small{}$70$}\tabularnewline
\hline 
{\small{}$\Delta\phi_{\mathrm{l^{+}},\mathrm{I}}+\Delta\phi_{\mathrm{l^{-}},\mathrm{I}}<$} & \multicolumn{5}{c|}{{\small{}$2$}}\tabularnewline
\hline 
{\small{}$M^{\tilde{\chi}_{1}^{\pm}}>$ {[}GeV{]}} & \multicolumn{5}{c|}{{\small{}$24$}}\tabularnewline
\hline 
{\small{}$\Delta\phi_{\mathrm{CM},\mathrm{I}}<$} & {\small{}$0.5$} & {\small{}$0.5$} & \multicolumn{3}{c|}{{\small{}$0.45$}}\tabularnewline
\hline 
{\small{}$\Delta\phi_{\mathrm{ISR},\mathrm{I}}>$} & {\small{}$3.12$} & {\small{}$3.10$} & {\small{}$3.06$} & {\small{}$3.0$5} & {\small{}$3.0$4}\tabularnewline
\hline 
{\small{}$\cos\theta>$ } & {\small{}$0.9$} & {\small{}$0.85$} & {\small{}$0.8$} & {\small{}$0.75$} & {\small{}$0.7$}\tabularnewline
\hline 
{\small{}$R_{\mathrm{ISR}}>$} & {\small{}$0.85,\;0.9$} & {\small{}$0.85,\;0.9$} & {\small{}$\begin{array}{c}
0.8,\;0.85\\
0.9
\end{array}$} & {\small{}$0.8,\;0.85$} & \foreignlanguage{british}{$\begin{array}{c}
0.75,\;0.8\\
0.85
\end{array}$}\tabularnewline
\hline 
\end{tabular}\caption{Selection criteria for signal regions in the analysis of chargino
pair production in final states with two leptons and missing transverse
energy. \label{tab:A-SignRegCCA}}
\end{table}
\begin{figure}[t]
\noindent \begin{centering}
\subfloat[\label{fig:CCN-1RISR}]{\includegraphics[height=5cm]{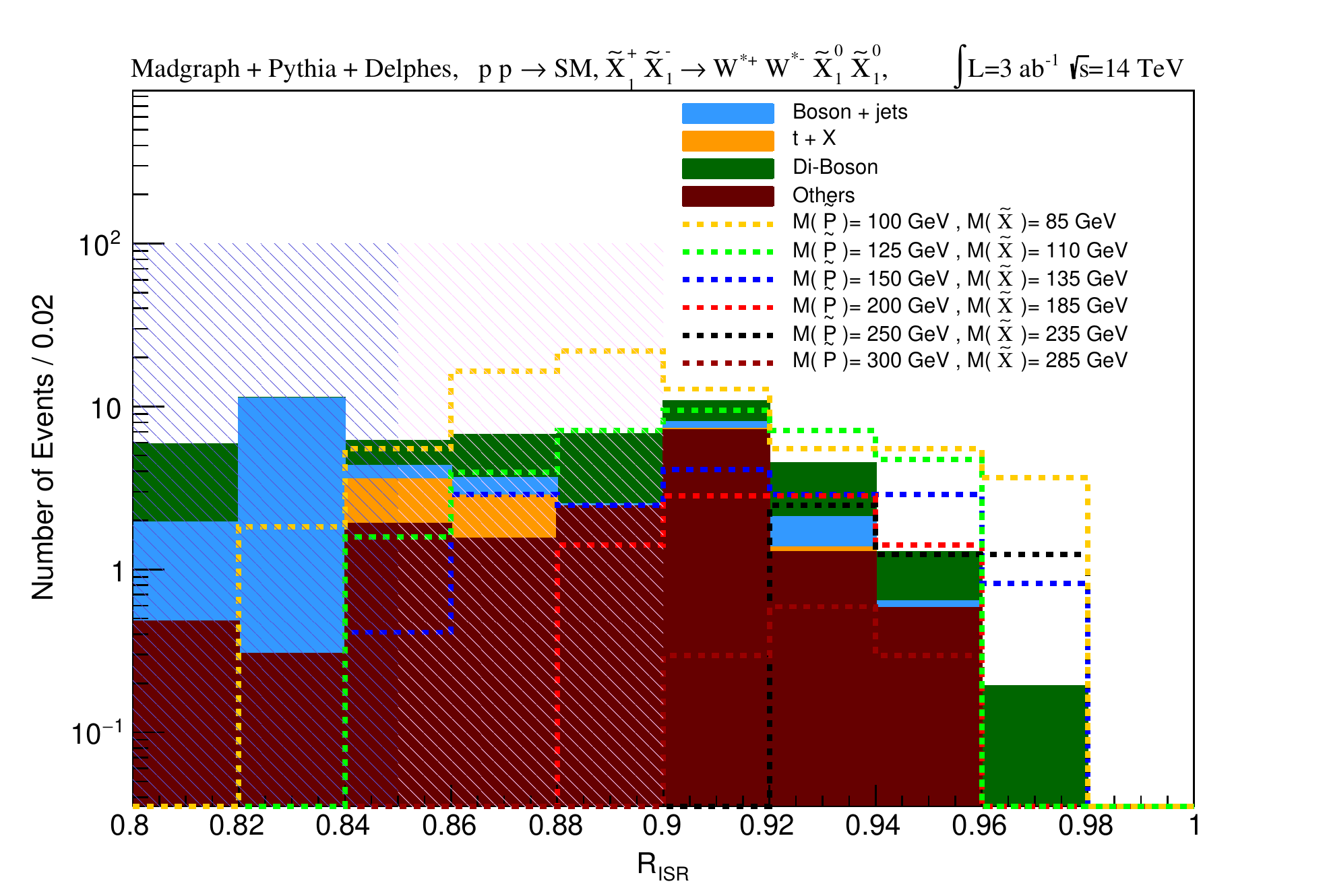}}
\subfloat[\label{fig:MCN-1CC}]{\includegraphics[height=5cm]{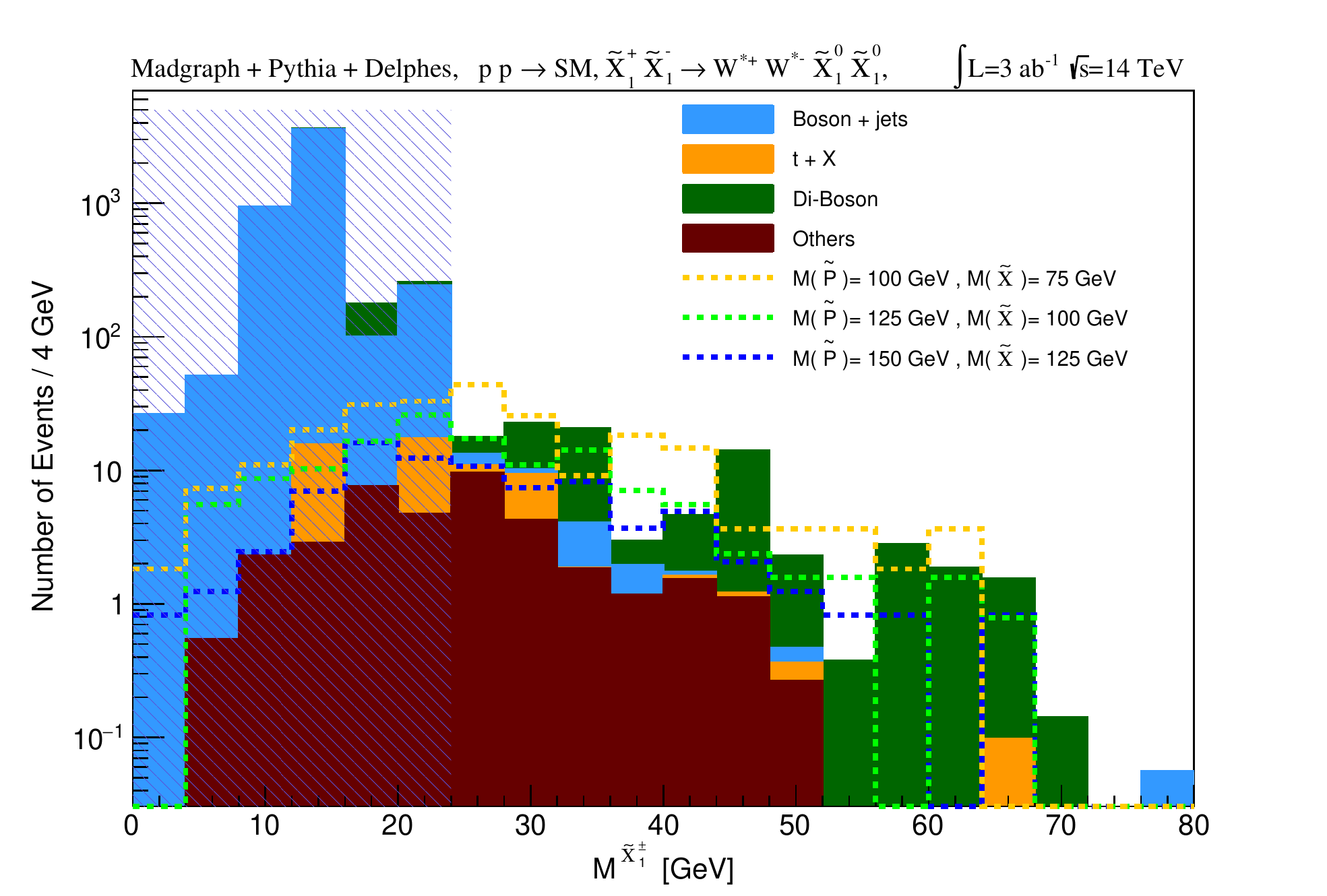}}
\par\end{centering}

\caption{The distributions of $R_{\mathrm{ISR}}$ for the signal and BG events
passing the N-1 selection criteria in Table \ref{tab:A-SignRegCCA}
column 1 (a) and of $M^{\mathrm{\tilde{\chi}^{\pm}}}$imposing
the requirements in column 2 with $R_{\mathrm{ISR}}>0.85$ (b)\label{fig:The-distributions-ofRISRMCA}.}
\end{figure}
\begin{figure}[t]
\noindent \centering{}\includegraphics[height=8cm]{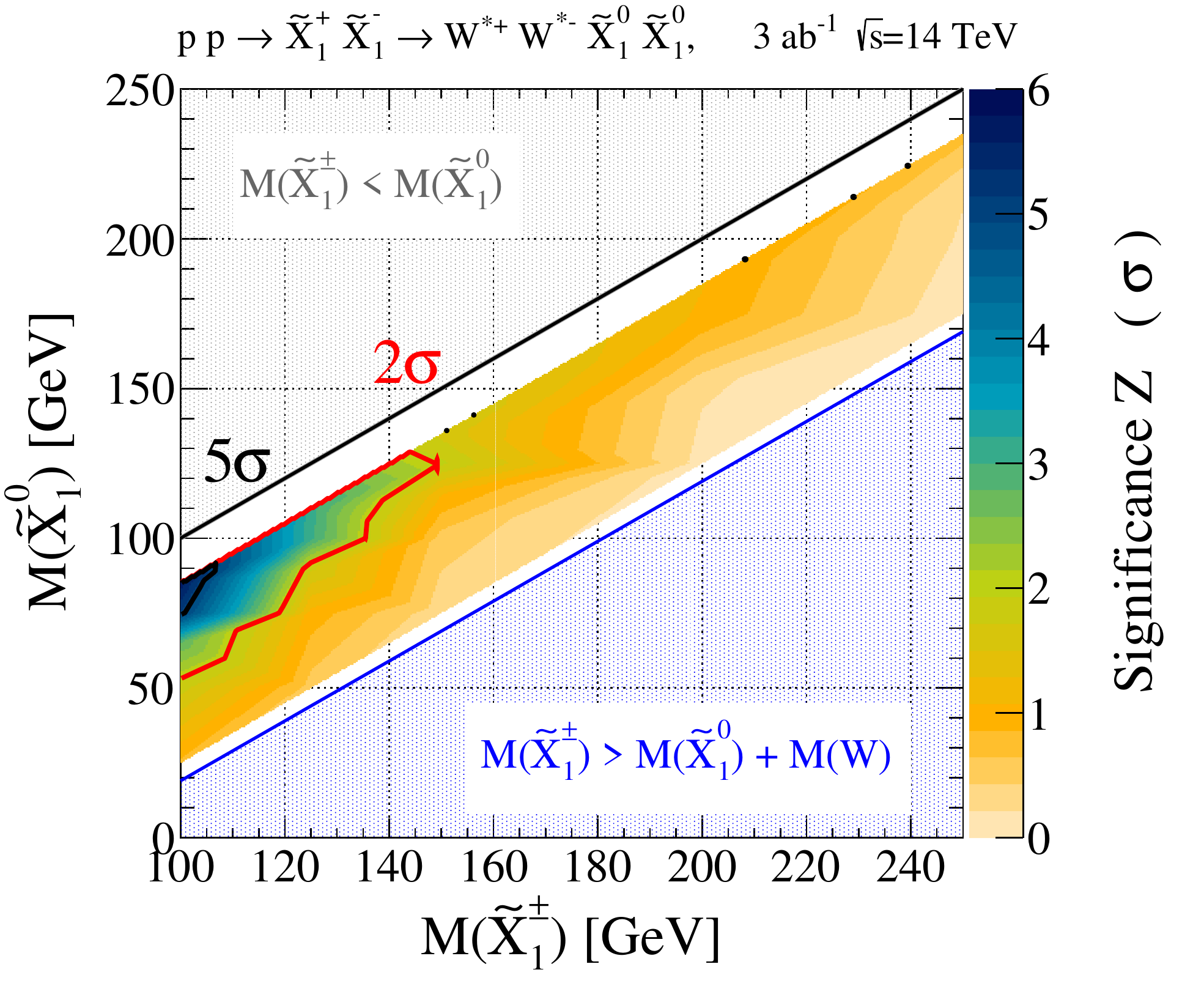}
\caption{Projected Z-value for chargino pair production in the compressed
region (15 GeV $\leq\Delta M\leq$ 75 GeV) at $\sqrt{s}=$ 14 TeV
assuming a systematic uncertainty of 20\% for the SM background
for an integrated luminosity of 3000 $\mathrm{fb}^{-1}$.\label{fig:ZValue_CC_3000}}
\end{figure}

Figure \ref{fig:The-distributions-ofRISRMCA} shows the distributions
of $R_{\mathrm{ISR}}$ and $M^{\mathrm{\tilde{\chi}^{\pm}}}$for SM
background and signal sample events passing the selection criteria
in Table \ref{tab:A-SignRegCCA}. For the lowest mass splitting the
requirement $R_{\mathrm{ISR}}>0.85$ is applied only for the sample
$M_{\tilde{\chi}_{1}^{\pm}}=100$ GeV, while for $\Delta M=25$ GeV
one demands this criterion for three samples ($M_{\tilde{\chi}_{1}^{\pm}}\leq150$
GeV).

The signal regions expressed by the selection criteria of the RJR
observables defined in Table \ref{tab:A-SignRegCCA} are applied to
calculate projected sensitivities for compressed spectra signal samples.
Figure \ref{fig:ZValue_CC_3000} shows the value of $Z_{Bi}$, the binomial score 
representing the significance of a given signal expressed in standard deviations 
in the presence of a background hypothesis, at $\sqrt{s}$=14 TeV for
an integrated luminosity of 3000 $\mathrm{fb}^{-1}$. One considers
a systematic uncertainty of 20\% for the overall Standard Model
background: a compromise between a large data sample projection (10
times the integrated luminosity of the associated chargino-neutralino
production analysis) and stringent selection criteria assumed to suppress
the background yields. 

Exploiting the RJR technique and with enough data collection one can
set limits for the compressed chargino pair production topology at
LHC14, with masses $\sim150$ GeV being excluded in the best
scenarios.

\section{Conclusions}

We have introduced an original approach to searches for compressed
electroweakinos based on the imposition of the decay trees as in Figures
\ref{fig:The-compressed-decay} and \ref{fig:The-compressed-treeCC}
for the interpretation of reconstructed events, using the Recursive
Jigsaw Reconstruction technique.

Putative wino-like chargino neutralinos could be discovered at LHC14
with masses $M_{\tilde{\chi}_{1}^{\pm}}=M_{\tilde{\chi}_{2}^{0}}$>150
GeV for a large portion of the samples investigated (15 GeV$\lesssim\Delta M\lesssim$50
GeV) assuming an integrated luminosity of 300 $\mathrm{fb}^{-1}$
and leveraging on only \textit{transverse} observables. The RJR technique
is sensitive to the extremely challenging chargino pair topology
scenarios in the compressed regime. A strategy based on several experimental
observables has been used to reduce the $W^{+}W^{-}$ and the
other main background yields due to the necessity of requiring jets
in the final state to be associated to the ISR-system. A potential
95\% confidence level exclusion limit can be obtained for an assumed data set 
of 3 $\mathrm{ab}^{-1}$ assuming a 20\% of systematic uncertainty
for sample spectra with $\Delta M\lesssim50$ GeV.

For both the topologies, the signal yields in the extreme compressed
scenarios can benefit from an improvement in the efficiencies of the
detector in the reconstruction of low transverse momentum leptons
($<10$ GeV). On the other hand, for large mass splittings ($\Delta M\apprge M_{Z}$
) the bulk analysis should be preferred to a compressed investigation,
while for intermediate scenarios $50\mathrm{\;GeV}\lesssim\Delta M\lesssim M_{Z}$
one can exploit the complementarity of observables based on a reconstruction
of the event with or without the ISR-system and include cases 
with vector bosons decaying hadronically.

The method is expected to have still more impact in the cases of final
state topologies with larger lepton multiplicity: pair production
of charginos and/or neutralinos with slepton mediated decays. The
RJR technique can be extended to these studies and to the pair production
of heavy neutralinos in final states with four leptons exploiting
the simplified tree in Figure \ref{fig:The-compressed-decay}, with
a simple modification in the assignment of the objects in the 
case of sleptons of the third generation. 

The results from the simplified models investigated in this work can
be partially reinterpreted assuming different compositions for the
electroweakinos. The method can be applied for higgsino-dominated
charginos and neutralinos, with the latter decaying via an off-shell Standard
Model Higgs boson, requiring two $b$-jets and one lepton in the V-system.
For chargino pair production with a mixed higgsino-wino nature,
one can re-weight the signal yields with the appropriate cross
sections: typically the contributions from off-shell charged Higgs
or other sparticles can be neglected since $M_{S},M_{H^{\pm}}\gg M_{W}$
in most SUSY models.

\bibliographystyle{JHEP}
\bibliography{marco}

\end{document}